\documentclass[%
 reprint,
 amsmath,amssymb,
 aps,
]{revtex4-2}

\usepackage{graphicx}
\usepackage{dcolumn}
\usepackage{bm}
\usepackage{array}
\usepackage{makecell}

\newcommand{\moire}{moir{\'e} }
\newcommand{\Moire}{Moir{\'e} }
\newcommand{\Kekule}{Kekul\'{e} }

\newcommand{\dIdVSG}{dI/dV(\ensuremath{V_s}, \ensuremath{V_g}) }

\newcommand{\filling}{\ensuremath{\nu}}
\newcommand{\Chern}{\ensuremath{\mathbb{C}}}
\newcommand{\EF}{\ensuremath{E_F}}

\newcommand{\WS}{WS\ensuremath{_2} }
\newcommand{\WSe}{WSe\ensuremath{_2} }
\newcommand{\MoTe}{MoTe\ensuremath{_2} }

\begin{document}

\preprint{APS/123-QED}

\title{A Microscopic Perspective on \Moire Materials}

\author{Kevin P. Nuckolls$^{1,2,\dagger}$}
\author{Ali Yazdani$^{1,\dagger}$}

\affiliation{%
 $^1$Joseph Henry Laboratories and Department of Physics, Princeton University, Princeton, NJ 08544, USA \\
 $^2$Department of Physics, Massachusetts Institute of Technology, Cambridge, MA, USA \\
 $^\dagger$Email: kpn@mit.edu, yazdani@princeton.edu
}%

\date{\today}

\begin{abstract}
\hspace{2mm} Contemporary quantum materials research is guided by themes of topology and of electronic correlations. A confluence of these two themes is engineered in ``\moire materials'', an emerging class of highly tunable, strongly correlated two-dimensional (2D) materials designed by the rotational or lattice misalignment of atomically thin crystals. In \moire materials, dominant Coulomb interactions among electrons give rise to collective electronic phases, often with robust topological properties. Identifying the mechanisms responsible for these exotic phases is fundamental to our understanding of strongly interacting quantum systems, and to our ability to engineer new material properties for potential future technological applications. In this Review, we highlight the contributions of local spectroscopic, thermodynamic, and electromagnetic probes to the budding field of \moire materials research. These techniques have not only identified many of the underlying mechanisms of the correlated insulators, generalized Wigner crystals, unconventional superconductors, \moire ferroelectrics, and topological orbital ferromagnets found in \moire materials, but they have also uncovered fragile quantum phases that have evaded spatially averaged global probes. Furthermore, we highlight recently developed local probe techniques, including local charge sensing and quantum interference probes, that have uncovered new physical observables in \moire materials.
\end{abstract}

\maketitle

\tableofcontents



\section{\label{sec:Intro}Introduction}
\subsection{Six Years of \Moire Materials}

\setlength{\columnsep}{0pt}
\setlength{\intextsep}{0pt}
\begin{figure*}
    \centering
    \includegraphics[width=\textwidth]{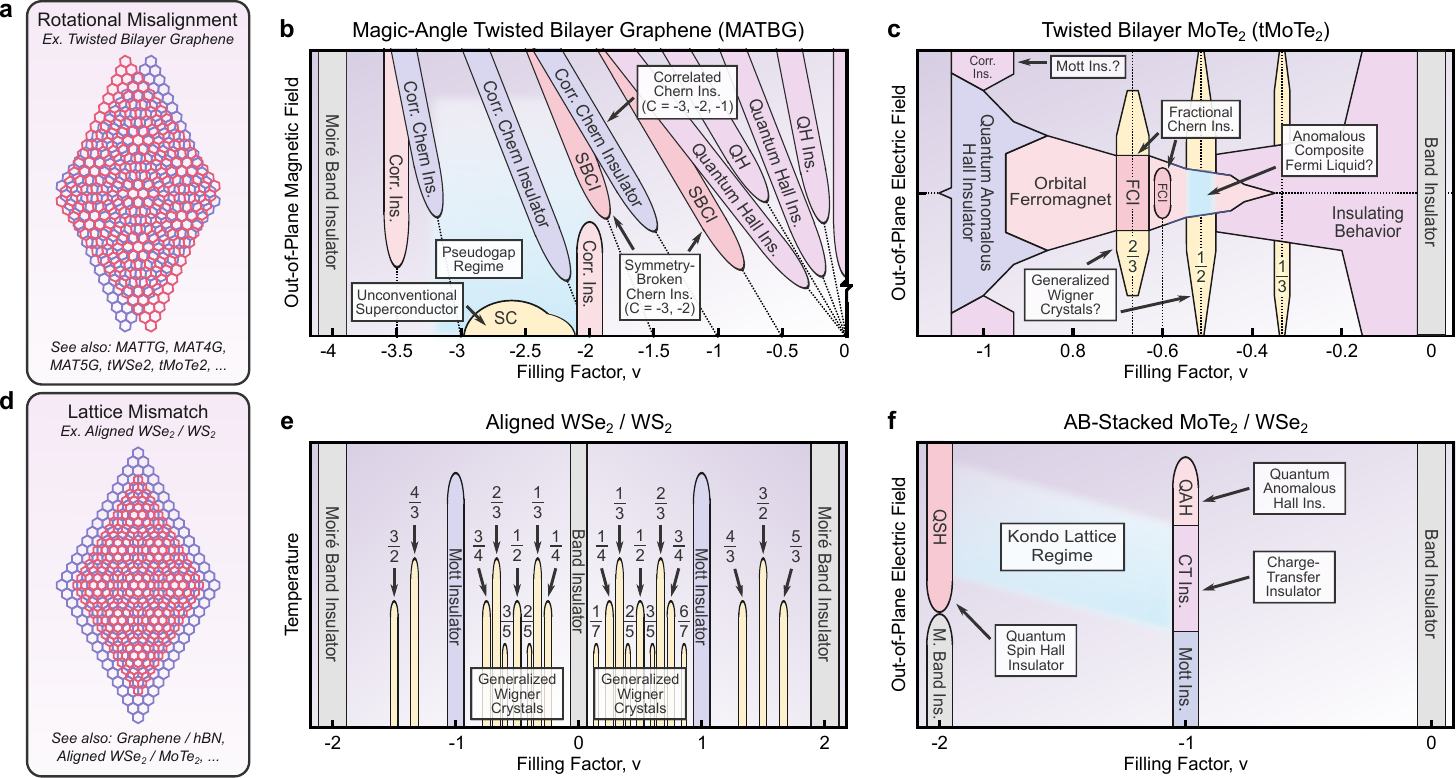}
    \caption{\label{fig:PhaseDiagrams} \textbf{Phase Diagrams of \Moire Materials.} \textbf{a,} Schematic diagram of a rotationally misaligned \moire material. When 2D materials with the same atomic lattice constants are twisted with respect to one another, a long-wavelength \moire interference pattern is produced. \textbf{b,} Schematic phase diagram of magic-angle twisted bilayer graphene (MATBG) as a function of out-of-plane magnetic field and filling ($\nu$), which hosts correlated insulators (Corr. Ins.), an unconventional superconductor (SC), a high-field pseudogap regime, quantum Hall insulators (QH Ins.), and field-stabilized correlated Chern insulators and symmetry-broken Chern insulators (SBCI). \textbf{c,} Schematic phase diagram of twisted bilayer \MoTe (tMoTe$_2$) as a function of out-of-plane electric field and $\nu$, which hosts a correlated insulators (likely Mott and generalized Wigner (GWC) phases at integer and fractional $\nu$, respectively), an integer quantum anomalous Hall insulator, fractional Chern insulator (FCI) phases showing fractional quantum anomalous Hall effects, and a possible anomalous composite Fermi liquid regime. \textbf{d,} Schematic diagram of a lattice-mismatched \moire material. When 2D materials with the different atomic lattice constants are aligned with one another, a long-wavelength \moire interference pattern is produced. \textbf{e,} Schematic phase diagram of aligned \WSe / \WS as a function of temperature and $\nu$, which hosts Mott insulators at integer $\nu$ and GWCs at fractional $\nu$. \textbf{f,} Schematic phase diagram of AB-stacking-aligned \MoTe / \WSe as a function of out-of-plane electric field and $\nu$, which hosts a Mott insulator (Mott Ins.), a charge-transfer insulator (CT Ins.), a quantum anomalous Hall insulator (QAH), a quantum spin Hall insulator (QSH), and a Kondo lattice regime.
    }
\end{figure*}

\hspace{2mm} Quantum materials research has experienced a veritable renaissance in roughly the past six years, with a surge of experimental reports of nearly every known electronic phase of matter and of several unique and unexpected phases, all found in an emerging class of highly tunable two-dimensional (2D) materials known as ``\moire materials'' \cite{balents2020superconductivity, andrei2021marvels, mak2022semiconductor}. \Moire materials are designed through either the rotational misalignment of identical 2D atomic crystals (Fig. \ref{fig:PhaseDiagrams}a) or the lattice-mismatch of dissimilar 2D atomic crystals (Fig. \ref{fig:PhaseDiagrams}d). Both of these incommensurability conditions cause long-wavelength ($\approx 10$ nm) interference patterns between the two constituent atomic lattices, forming an enlarged \moire superlattice that generically hosts flat electronic bands, which are highly conducive to correlated, collective phases of matter.

\hspace{2mm} Today, there are roughly a dozen distinct \moire material platforms designed by this prescription that have been the subject of immense theoretical and experimental research efforts. The rotational misalignment design motif describes a wide variety of \moire homobilayers, including twisted monolayer graphene (ex. magic-angle twisted bilayer graphene \cite{cao2018correlated,cao2018unconventional,yankowitz2019tuning,lu2019superconductors}, Fig. \ref{fig:PhaseDiagrams}b), twisted multilayer graphene (ex. twisted monolayer-bilayer graphene \cite{polshyn2020electrical,chen2021electrically,polshyn2022topological}, twisted double bilayer graphene \cite{liu2020tunable,cao2020tunable,shen2020correlated}) and twisted transition metal dichalcogenides (ex. twisted bilayer \WSe \cite{wang2020correlated,ghiotto2021quantum,xu2022tunable}, \WS \cite{li2022mapping,li2024imaging}, WTe$_2$ \cite{wang2022one,yu2023evidence} and \MoTe \cite{anderson2023programming,cai2023signatures,zeng2023integer}; Fig. \ref{fig:PhaseDiagrams}c), and has further expanded to include multilayer structures with multiple twisted interfaces (ex. magic-angle twisted trilayer / quadralayer / pentalayer graphene \cite{park2021tunable, hao2021electric, park2022robust, zhang2022promotion}, \moire quasicrystalline trilayer graphene \cite{uri2023superconductivity}, helical trilayer graphene \cite{xia2023helical}). The lattice mismatch motif describes a series of aligned heterobilayer transition metal dichalcogenides (ex. aligned \WSe / \WS \cite{tang2020simulation,regan2020mott}, Fig. \ref{fig:PhaseDiagrams}e; aligned \MoTe / \WSe \cite{li2021quantum,zhao2022realization,zhao2023gate}, Fig. \ref{fig:PhaseDiagrams}f) and aligned graphene / hexagonal boron nitride (hBN) heterostructures (ex. monolayer / bilayer graphene aligned to hBN \cite{yankowitz2012emergence,dean2013Hofstadter,hunt2013massive,ponomarenko2013cloning,yankowitz2019van}). Furthermore, many more \moire material platforms have been predicted to host exotic correlated electronic states, and are awaiting experimental realization \cite{khalaf2019magic, ledwith2022family, lian2020flat, zhang20214, shi2021moire, can2021high, popov2023magic, yang2023multi, kim2023controllable}.

\hspace{2mm} \Moire materials sit naturally at the confluence of two influential paradigms of quantum materials research: electronic correlation effects and topologically protected properties. Electronic correlation effects arise from many-body interactions among electrons that produce complex and collective phenomena. In \moire materials, emergent flat \moire bands quench the kinetic energy of the dispersive electronic bands found in natural 2D materials by nearly 3 orders of magnitude \cite{morell2010flat,bistritzer2011moire}, causing Coulomb interaction energies to dominate the system’s dynamics. These settings are conducive to correlated phases, the nature of which depends critically on properties unique to each \moire material platform. Topologically protected properties describe the unchangeable properties of materials derived from geometric qualities of the material’s energy bands. The low-energy band structures of many commonly used 2D materials in \moire materials (ex. graphene, transition metal dichalcogenides (TMDs)) are described by the Dirac physics associated with the hexagonal atomic lattice motif. Furthermore, \moire materials made from TMD layers host intrinsic spin-orbit coupling conducive to topological properties.

\hspace{2mm} Because these effects co-exist in \moire materials, correlations and topology come together to produce novel electronic phases, some of which were previously unrealizable within either paradigm individually. The correlation-driven topological phases found in many \moire materials are notable examples of this synergy. Typically, topological phases are derived from single-particle effects, and are most commonly found in materials with strong spin-orbit coupling \cite{hasan2010colloquium, qi2011topological, bernevig2013topological, sato2017topological, burkov2016topological, yan2017topological, armitage2018weyl, tokura2019magnetic}. In contrast, correlations in \moire materials can break symmetries that protect previously inaccessible sources of topological Berry curvature in 2D materials, producing new and unexpected correlated topological phases \cite{sharpe2019emergent,chen2020tunable,serlin2020intrinsic, nuckolls2020strongly,wu2021chern,saito2021hofstadter,das2021symmetry, choi2021correlation,park2021flavour,yu2022correlated}, some without single-particle analogues (ex. fractional Chern insulators) \cite{xie2021fractional,cai2023signatures,zeng2023integer, park2023observation, xu2023observation}.


\subsection{Rich, Tunable Phase Diagrams of \Moire Materials}

\hspace{2mm} We begin with a brief summary of the electronic landscape of a few of the most studied \moire materials to-date. Fig. \ref{fig:PhaseDiagrams} shows a series of schematic phase diagrams, each the culmination of the collective efforts of dozens of research groups around the world. Fig. \ref{fig:PhaseDiagrams}b represents the phase diagram of magic-angle twisted bilayer graphene (MATBG) as a function of out-of-plane magnetic field and filling ($\nu$). At zero magnetic field, MATBG hosts a correlated insulator near \filling \: $= -2$ \cite{cao2018correlated} and an unconventional superconductor upon hole-doping this insulator \cite{cao2018unconventional}. The microscopic nature of these phases will be the focus of Sections \ref{sec:Corr_Graphene} and \ref{sec:SC}, respectively. These zero-field ground states are quenched in a magnetic field, giving rise to a high-field pseudogap regime \cite{oh2021evidence}, a precursor to superconductivity, and a sequence of integer \cite{nuckolls2020strongly, wu2021chern,saito2021hofstadter,das2021symmetry,choi2021correlation,park2021flavour} and symmetry-broken Chern insulators \cite{saito2021hofstadter,yu2022correlated}, which will be the focus of Section \ref{sec:Topo}.

\hspace{2mm} Fig. \ref{fig:PhaseDiagrams}c represents the phase diagram of twisted bilayer \MoTe (tMoTe$_2$) as a function of out-of-plane electric field and filling. Unlike MATBG, tMoTe$_2$ admits an electric-field tunable flat band structure derived from the inequivalent stacking regions of its uniquely hexagonal \moire superlattice. tMoTe$_2$ displays spontaneous orbital ferromagnetism within a wide range of fillings \cite{anderson2023programming}, which additionally stabilizes a quantum anomalous Hall (QAH) insulator (i.e. Chern insulator at zero magnetic field) at filling \filling \: $= -1$ and fractional Chern insulators (FCIs) showing fractional quantum anomalous Hall effects at fillings \filling \: $= -2/3, -3/5$ \cite{cai2023signatures,zeng2023integer, park2023observation, xu2023observation}. A series of topological phase transitions are induced with increasing electric field, resulting in correlated insulating phases at these fillings. These are likely Mott and generalized Wigner crystal (GWC) states \cite{regan2020mott, tang2020simulation}, although verification of this hypothesis will require further investigation using local imaging techniques. Finally, a compressible state with a large Hall resistance of 2h/e$^2$ appears near filling \filling \: $-1/2$ \cite{park2023observation}, possibly an anomalous composite Fermi liquid regime reminiscent of the half filling state of the lowest Landau level \cite{goldman2023zero, dong2023composite}.

\hspace{2mm} Fig. \ref{fig:PhaseDiagrams}e represents the phase diagram of aligned \WSe / \WS as a function of temperature and filling. The \moire superlattice potential of heterobilayer \moire materials made from TMDs is deeper than those of twisted graphene-based \moire materials \cite{shabani2021deep}, which favors charge-ordered states where electrons / holes are trapped near high-symmetry-stacking locations of the \moire superlattice. In aligned \WSe / \WS, charge-ordered Mott insulators are observed near \filling \: $= \pm 1$ \cite{tang2020simulation,regan2020mott}, and GWC phases (i.e. commensurate filling electron solid phases) are observed near roughly two dozen fractional fillings \cite{regan2020mott,xu2020correlated,huang2021correlated}. These charge-ordered phases will be the focus of Section \ref{sec:Corr_TMD}.

\hspace{2mm} Finally, Fig. \ref{fig:PhaseDiagrams}f represents the phase diagram of AB-stacking-aligned \MoTe / \WSe as a function of out-of-plane electric / displacement field and filling. Here, ``AB-stacking'' refers to a near $60^\circ$ twist angle between \MoTe and \WSe layers, distinct from ``AA-stacking'' (twisted near $0^\circ$) because of the threefold rotational symmetry of the TMD layers. Near \filling \: $= -1$, AB-stacked \MoTe / \WSe hosts a correlated insulator, the nature of which depends on the magnitude of the electric field \cite{li2021quantum,zhao2023gate}. The Mott insulator is doped into a metallic state with quantum oscillations associated with the bands of both \MoTe and WSe$_2$. The charge-transfer insulator (CT Ins.) is doped into a metal with quantum oscillations only associated with the \WSe layer, while the \MoTe layer remains insulating and hosts a lattice of local moments. The coupling between these layers produces a tunable \moire Kondo lattice regime \cite{zhao2023gate}, persisting to \filling \: $= -2$, where a quantum spin Hall (QSH) insulator appears over a broad range of displacement field \cite{zhao2022realization}. Further increasing the displacement field at \filling \: $= -1$ produces a Chern $+1$ insulator with a quantized anomalous Hall response \cite{li2021quantum}.


\subsection{Capabilities of Local Probe Techniques}

\newcommand\picSize{0.36}

\setlength{\columnsep}{0pt}
\setlength{\intextsep}{0pt}

\begin{table*}
  \centering
  \begin{tabular}{ | p{\picSize\textwidth} | p{0.3\textwidth} | p{0.3\textwidth} | }
    \hline
    
    \begin{minipage}[c][7ex][c]{\picSize\textwidth}
        Local Probe Technique
    \end{minipage}
    &
    \begin{minipage}{.3\textwidth}
        Working Principle, Capabilities, and Limitations
    \end{minipage}
    &
    \begin{minipage}{.3\textwidth}
        Experimental Insights
    \end{minipage}
    \\ \hline
    \begin{minipage}{\picSize\textwidth}
      \includegraphics[width=\linewidth]{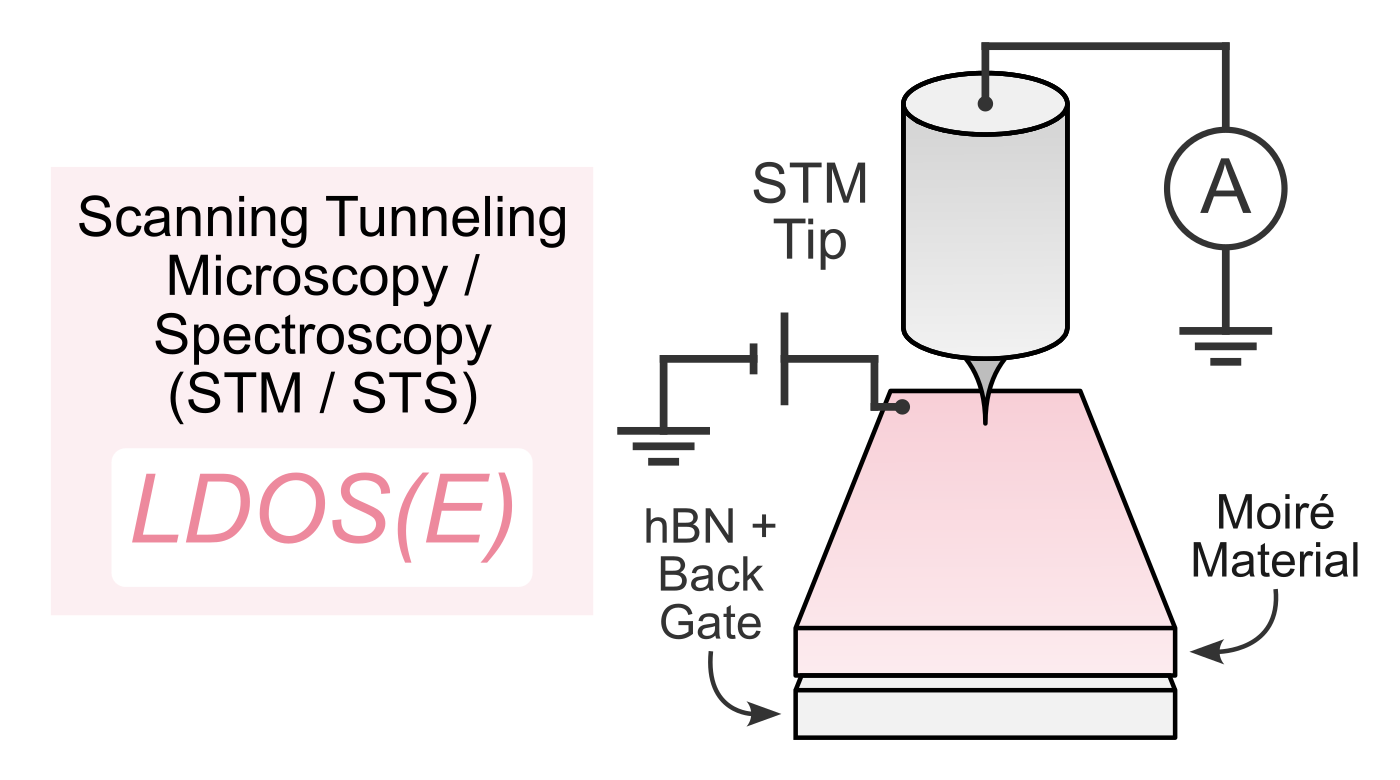}
    \end{minipage}
    &
    \begin{minipage}{.3\textwidth}
    \raggedright
    $\boldsymbol{\cdot}$ Voltage bias the sample and measure the quantum tunneling current through the tip-sample junction. \\
    
    $\boldsymbol{\cdot}$ Measures local density of states (LDOS $\propto$ $\frac{dI}{dV}$) with $\approx$ $50$ $\mu$eV energy resolution and atomic-scale spatial resolution. Limited to back-gated device geometries.
    \end{minipage}
    &
    \begin{minipage}{.3\textwidth}
    \raggedright
    $\boldsymbol{\cdot}$ Spectroscopically measure flat bands and their one-particle excitations (i.e. spectral function) \\
    
    $\boldsymbol{\cdot}$ Density-dependent spectroscopy of energetically gapped phases (i.e. insulating, superconducting, topological phases) \\

    $\boldsymbol{\cdot}$ Wavefunction mapping of correlated phases in ultra-clean materials
    \end{minipage}
    \\ \hline
    \begin{minipage}{\picSize\textwidth}
      \includegraphics[width=\linewidth]{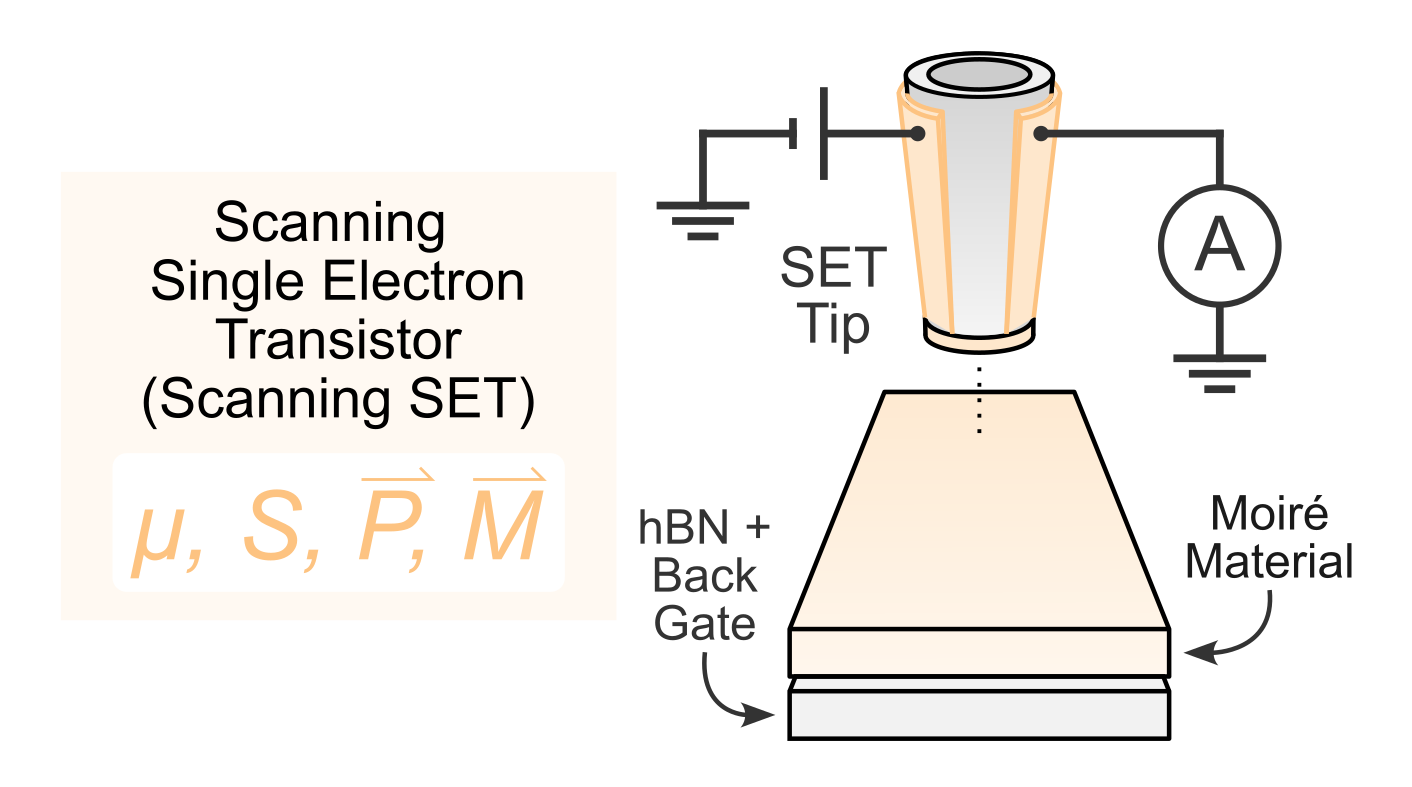}
    \end{minipage}
    &
    \begin{minipage}{.3\textwidth}
    \raggedright
    $\boldsymbol{\cdot}$ Measure the 2-terminal conductance of the tip SET, using a local gate to maintain voltage sensitivity. \\
     
    $\boldsymbol{\cdot}$ Measures electronic compressibility ($\frac{dn}{d\mu}$) with $\approx$ $100$ nm spatial resolution and $\approx$ 10$^{-5}$ $\mu$V/$\sqrt{}$Hz voltage sensitivity. Limited to back-gated device geometries (but compatible with hBN encapsulation).
    \end{minipage}
    &
    \begin{minipage}{.3\textwidth}
    \raggedright
    $\boldsymbol{\cdot}$ Measure the chemical potential ($\mu$) and electronic compressibility ($\frac{dn}{d\mu}$) of many-body phases \\

    $\boldsymbol{\cdot}$ Identify thermodynamic phase transitions between correlated phases \\

    $\boldsymbol{\cdot}$ Measure the entropy (S), electric polarization ($\vec{P}$), and magnetization ($\vec{M}$) of correlated phases via T-, $\vec{E}$-, and $\vec{B}$-dependence, respectively
    \end{minipage}
    \\ \hline
    \begin{minipage}{\picSize\textwidth}
      \includegraphics[width=\linewidth]{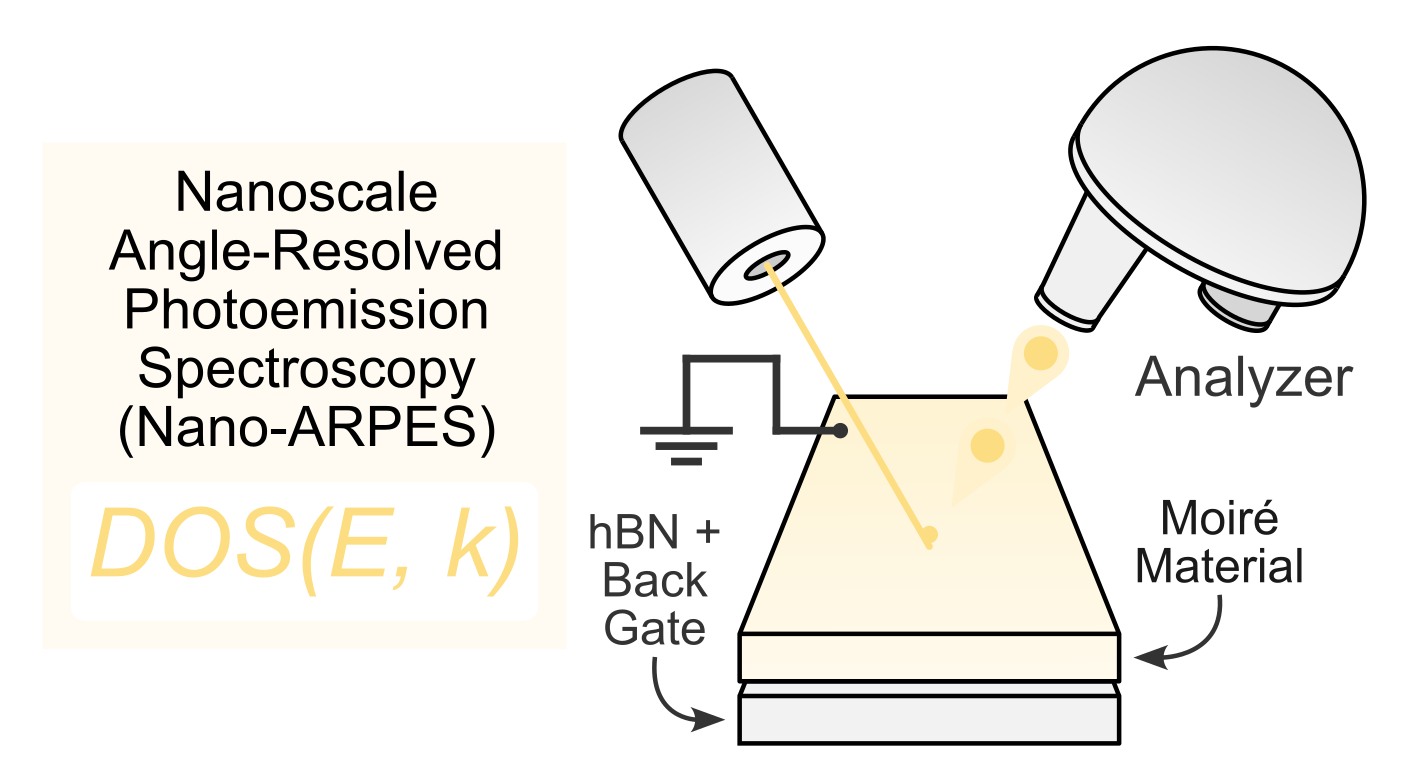}
    \end{minipage}
    &
    \begin{minipage}{.3\textwidth}
    \raggedright
    $\boldsymbol{\cdot}$ Shine monochromatic light on the sample and measure the energy and momentum of ejected electrons. \\

    $\boldsymbol{\cdot}$ Measures energy-momentum dispersions (DOS(E,k)) with $\approx$ $20$ meV energy resolution and sub-$\mu$m spatial resolution. Limited to back-gated device geometries and elevated temperatures (T $>$ $20$ K).
    \end{minipage}
    &
    \begin{minipage}{.3\textwidth}
    \raggedright
    $\boldsymbol{\cdot}$ Spectroscopically visualize the dispersion characteristics of the flat bands themselves \\

    $\boldsymbol{\cdot}$ Identify collective excitations that strongly couple to electronic degrees of freedom (ex. phonons, plasmons, magnons, etc.)
    \end{minipage}
    \\ \hline
    \begin{minipage}{\picSize\textwidth}
      \includegraphics[width=\linewidth]{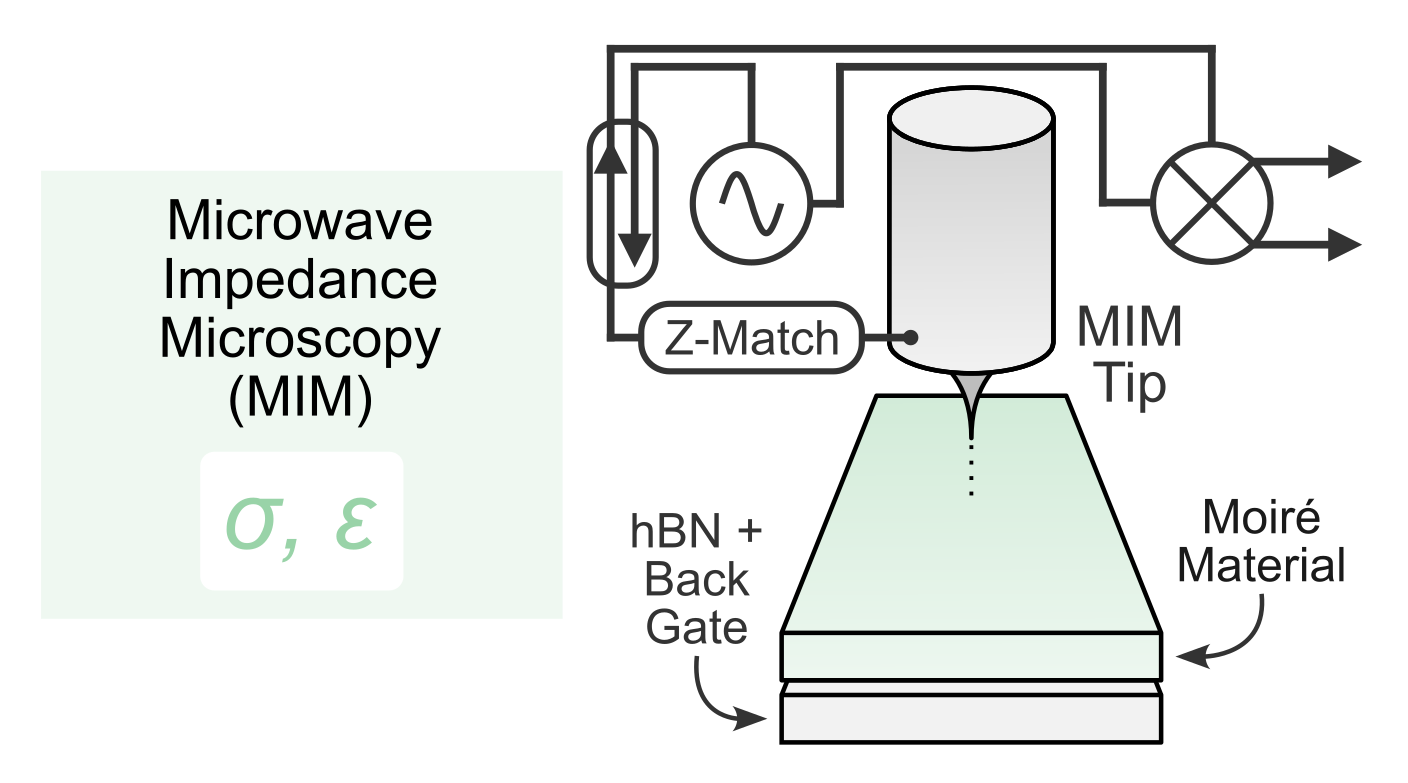}
    \end{minipage}
    &
    \begin{minipage}{.3\textwidth}
    \raggedright
    $\boldsymbol{\cdot}$ Drive sample locally with microwave radiation and measure the real / imaginary parts of the reflected signal.
    
    $\boldsymbol{\cdot}$ Measures local electrical conductivity and permittivity with $\approx$ $50$ nm spatial resolution. Limited to back-gated device geometries (but compatible with hBN encapsulation).
    \end{minipage}
    &
    \begin{minipage}{.3\textwidth}
    \raggedright
    $\boldsymbol{\cdot}$ Sensitive to correlated insulating phases that are fragile and difficult to stabilize homogeneously over micron-sized devices.
    \end{minipage}
    \\ \hline
    \begin{minipage}{\picSize\textwidth}
      \includegraphics[width=\linewidth]{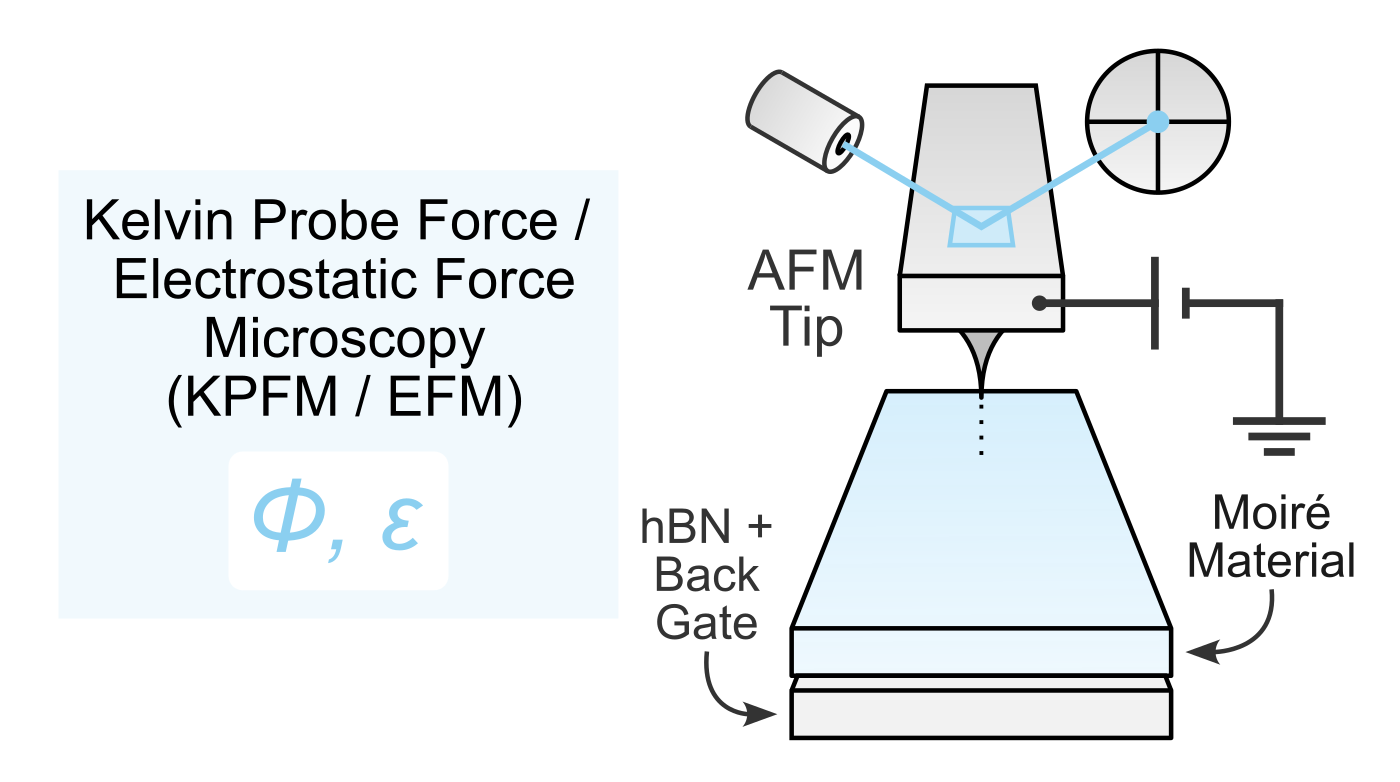}
    \end{minipage}
    &
    \begin{minipage}{.3\textwidth}
    \raggedright
     $\boldsymbol{\cdot}$ Voltage bias the tip and measure either the electrostatic force profile of the sample directly (dc-EFM) or the voltage necessary to cancel the electrostatic force signal (KPFM).

    $\boldsymbol{\cdot}$ Measures variations in a material's surface potential with as high as $\approx$ $1$ nm spatial resolution. Mostly limited to elevated temperatures (T $>$ $4$ K).
    \end{minipage}
    &
    \begin{minipage}{.3\textwidth}
    \raggedright
    $\boldsymbol{\cdot}$ Sensitive to inversion symmetry broken phases that admit electrostatic responses

    $\boldsymbol{\cdot}$ Identifies polar / ferroelectric phases

    $\boldsymbol{\cdot}$ Elucidates the domain structure and electric domain-switching mechanisms ferroelectric phases (and can control switching using the tip's electric field)
    \end{minipage}
    \\ \hline
    \begin{minipage}{\picSize\textwidth}
      \includegraphics[width=\linewidth]{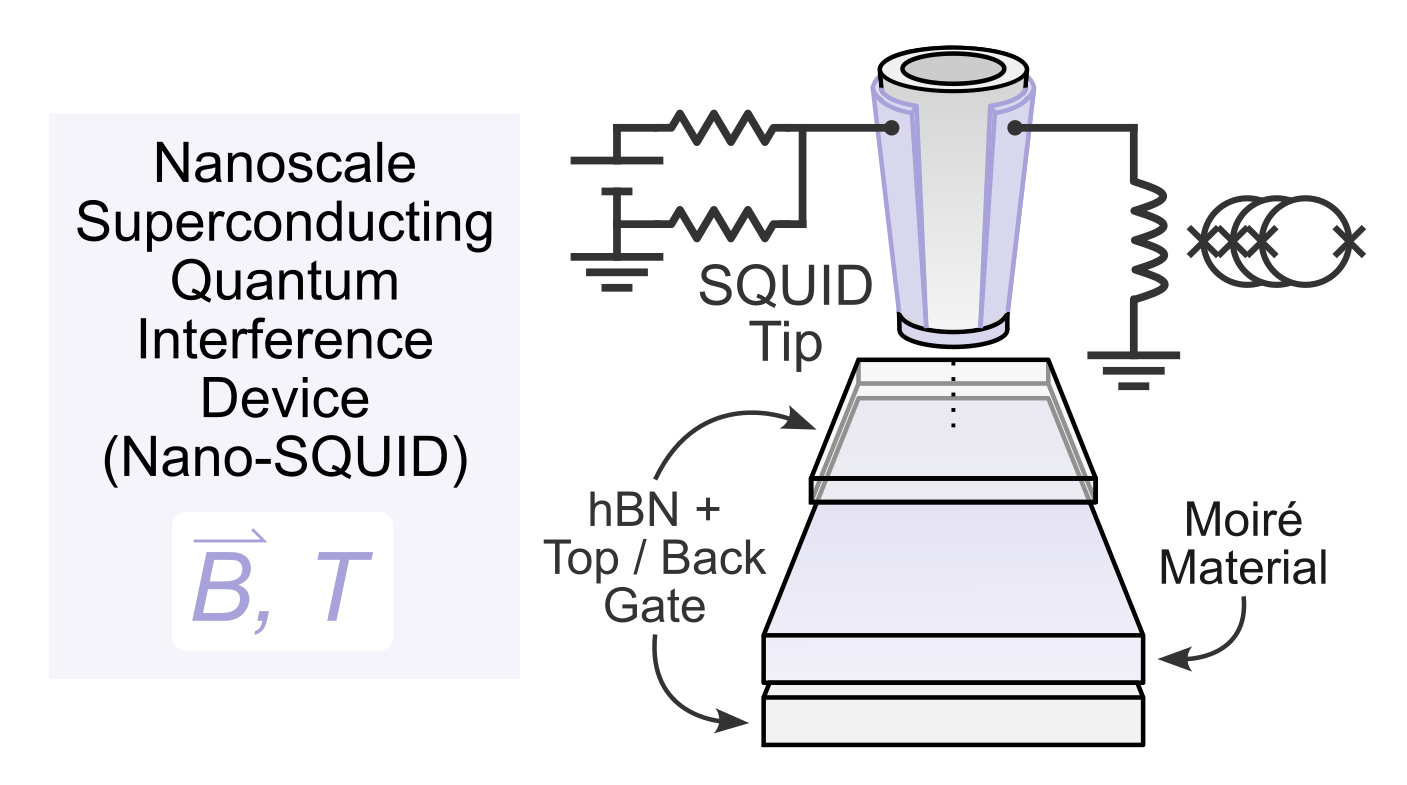}
    \end{minipage}
    &
    \begin{minipage}{.3\textwidth}
    \raggedright
    $\boldsymbol{\cdot}$ Voltage bias and measure the current through the tip's SQUID via a series SQUID array amplifier.

    $\boldsymbol{\cdot}$ Measures local B-field strength with submicron spatial resolution and magnetic field sensitivity of $\approx$ $15$ nT/$\sqrt{}$Hz at certain B-field values. Capable of probing dual-gated encapsulated devices.
    \end{minipage}
    &
    \begin{minipage}{.3\textwidth}
    \raggedright
    $\boldsymbol{\cdot}$ Sensitive to time-reversal symmetry broken phases that admit magnetostatic responses

    $\boldsymbol{\cdot}$ Elucidates the domain structure and domain-switching mechanisms magnetic phases, and relate this to in situ global transport measurements
    
    $\boldsymbol{\cdot}$ Images thermodynamic quantum oscillations to reconstruct local band structure features
    \end{minipage}
    \\ \hline
  \end{tabular}
  \caption{\label{table:LocalProbes} \textbf{Local Probe Techniques for Studying \Moire Materials.} }\label{tbl:myLboro}
\end{table*}

\hspace{2mm} Electrical transport measurements have revealed many phases in \moire materials, offering several advantages over competing techniques. They can identify insulating phases by their activated ``Arrhenius'' behavior, topological phases by their hysteretic Hall response, and superconducting phases by their vanishing resistance and associated Fraunhofer oscillations (although such signatures are, alone, not definitive). Additionally, transport measurements are highly compatible with the dual-gated device structures needed to access the full parametric phase space of electric-field tunable \moire materials.

\hspace{2mm} Following up on landmark transport experiments, however, local probe techniques (Table \ref{table:LocalProbes}) have deepened our understanding of these complex phase diagram and of the electronic mechanisms at play in \moire materials. These techniques include a suite of local spectroscopic probes that include scanning tunneling microscopy / spectroscopy (STM / STS) and micro- / nano-scale angle-resolved photoemission spectroscopy ($\mu$ARPES / nanoARPES), thermodynamic probes that include scanning single-electron transistor (scanning SET), and electromagnetic probes that include scanning nano-scale superconducting quantum interference device (nano-SQUID), scanning nitrogen-vacancy magnetometry (scanning NV), scanning nearfield optical microscopy (SNOM), microwave impedance microscopy (MIM), Kelvin probe force microscopy (KPFM), and electrostatic force microscopy (EFM). Each probe has offered a distinctive and complementary experimental perspective for distinguishing complex quantum phases in this class of materials.

\hspace{2mm} Spectroscopic probes like STM / STS and $\mu$ARPES / nanoARPES give experimentalists energy resolution to visualize the dispersion characteristics of \moire flat bands \cite{lisi2021observation, utama2021visualization}, to spectroscopically probe electronic excitations that distinguish many-body states \cite{wong2020cascade}, to identify collective excitations that strongly couple to electronic degrees of freedom \cite{chen2023strong}, and to quantify subtle effects of orbital magnetism through Landau-level spectroscopy \cite{slot2023quantum}. Moreover, the high energy-resolution of millikelvin-temperature STS experiments gives distinct insights into the nature of energetically gapped correlated phases, including correlated insulating, superconducting, and topological phases \cite{nuckolls2020strongly, choi2021correlation, oh2021evidence, kim2022evidence, nuckolls2023quantum, kim2023imaging}.

\hspace{2mm} As a thermodynamic probe, scanning SET gives access to the chemical potential and the electronic compressibility of \moire materials \cite{zondiner2020cascade, yu2022correlated, pierce2021unconventional,xie2021fractional}. This technique is singular in its ability to identify thermodynamic phase transitions between correlated phases, and to do so in microscopic ($\approx$ $0.01$ $\mu$m$^2$) regions of devices where fragile quantum phases can be stabilized. In addition, by performing such measurements as a function of temperature, out-of-plane electric field, or out-of-plane magnetic field (in conjunction with Maxwell’s relations), researchers gain access to the entropy, the electric polarization, or the magnetization of the system, respectively \cite{rozen2021entropic}. Such information can not only complement transport and spectroscopic experiments, but also provides insights into the nature of complex correlated phases that are otherwise inaccessible.

\hspace{2mm} Lastly, many electromagnetic probes are used to investigate correlated phases that admit broken inversion-symmetry and broken time-reversal symmetry, which produce measurable electrostatic and magnetostatic responses, respectively. Nano-SQUID techniques have characterized the magnetic response of orbital magnetic phases ubiquitously found in \moire materials \cite{tschirhart2021imaging, grover2022chern}. Scanning NV magnetometry has identified \moire magnetism, spatially modulating spin configurations, in twisted CrI$_3$ \cite{song2021direct, huang2023revealing}. KPFM, EFM, and SNOM measurements have identified ferroelectric phases and distinctive lattice dynamics occurring in stacking-engineered polar \moire materials \cite{woods2021charge, vizner2021interfacial, weston2022interfacial, deb2022cumulative, moore2021nanoscale}. Overall, the spatial resolution of these techniques provides unique insights into the domain structure and electric domain-switching mechanisms of magnetic and ferroelectric phases in \moire materials.


\subsection{Advantages of Local Probe Techniques}

\hspace{2mm} Broadly, local probe techniques have been successful in the budding field of \moire materials research because they give us access to the physical observables most relevant to identifying and distinguishing quantum phases of matter, often at the micrometer- or even the nanometer-scale. Moreover, local probe techniques are robust against the extrinsic inhomogeneity that has plagued early studies in this field \cite{lau2022reproducibility}. Experimental studies of \moire materials often struggle to address severe sample-to-sample variations, stemming from the energetically competitive nature of a manifold of ground states in these systems. Such states are extremely sensitive to the presence of classic types of sample disorder, like charge inhomogeneity, but also the presence of new types of sample disorder particular to \moire materials, like twist angle gradients or interlayer heterostrain \cite{uri2020mapping}. In these settings, local probe techniques present a robust perspective to understanding these correlated ground states because they can often not only quantify the disorder landscape of a particular device location (ex. strain and twist angle), but also relate this characterization to the electronic phenomenology measured in that same sample location.

\hspace{2mm} Finally, local probe techniques are particularly informative in the presence of intrinsic (i.e. electronic) inhomogeneity. This advantage helps distinguish the microscopic mechanisms responsible for electronic phases, and emphasizes the importance of spatial resolution to local probe techniques. Consider the situation where electrical transport experiments measure an insulating phase in a 2D material. For that same state, local probes like the STM can use the spatial distribution of the local density of states to identify the nature of this insulating phase (ex. a charge-density wave state in NbSe$_2$ \cite{xi2015strongly, ugeda2016characterization}, a Mott insulating state in TaSe$_2$ \cite{chen2020strong}, or an incommensurate \Kekule spiral state in MATBG \cite{nuckolls2023quantum, caluguaru2022spectroscopy, hong2022detecting}). Recent innovations in van der Waals device geometries have further equipped local probe techniques in their abilities to measure new physical observables, like the real-space distribution of charge, which has enabled the direct visualization of new electronic order (ex. a series of GWC and stripe phases in aligned \WSe / \WS) \cite{li2021imaging, li2022mapping}.

\hspace{2mm} In this Review, we highlight many important discoveries of exotic quantum phases in \moire materials. We emphasis how researchers have utilized the aforementioned advantages of local probe techniques to uncover the underlying electronic mechanisms responsible for the unprecedentedly tunable phase diagrams of \moire materials.

\section{\label{sec:FlatBands}Flat Electronic Bands}
\subsection{Flat Electronic Bands}

\setlength{\columnsep}{0pt}
\setlength{\intextsep}{0pt}
\begin{figure*}
    \centering
    \includegraphics[width=\textwidth]{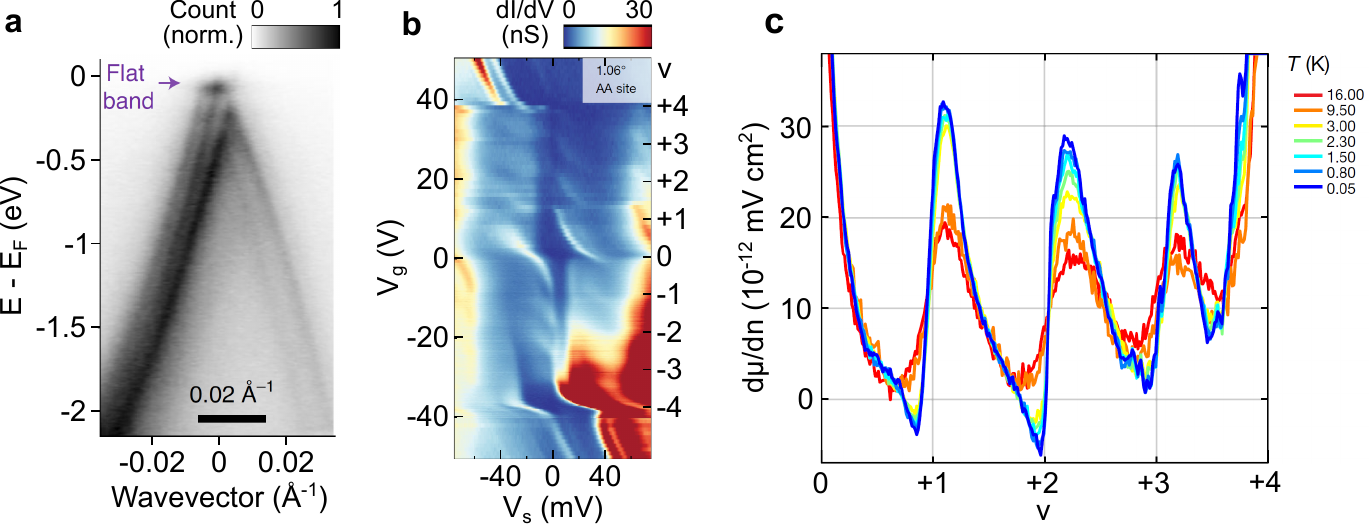}
    \caption{\label{fig:FlatBands} \textbf{Flat Electronic Bands and Cascades of Electronic Transitions.} \textbf{a,} Energy-momentum band dispersion of magic-angle twisted bilayer graphene (MATBG) around the K point of the Brillouin zone, as measured in angle-resolved photoemission spectroscopy. A purple arrow marks the lower flat band of MATBG. \textbf{b,} Differential conductance dI/dV(V$_s$, V$_g$) of the cascade of transitions in MATBG, measured in scanning tunneling spectroscopy. A sudden reorganization of the low-energy excitations of MATBG occurs near each integer filling of the \moire flat bands. \textbf{c,} Inverse compressibility d$\mu$/dn($\nu$) of the cascade of transitions in MATBG, measured by the scanning single-electron transistor. The sawtooth-shaped signal is indicative of resets of the chemical potential near integer fillings. Reprinted from \cite{utama2021visualization,wong2020cascade,zondiner2020cascade} with minor adaptations for style and formatting purposes.
    }
\end{figure*}

\hspace{2mm} \Moire materials have been investigated for several decades, prior to the observation of correlated insulators and superconductivity in MATBG \cite{cao2018correlated,cao2018unconventional}. In fact, the first observations of a \moire superlattice appeared in early STM studies of the surface of naturally turbostratic graphite \cite{kuwabara1990anomalous, rong1993electronic, xhie1993giant}. Here, STM topographic images show a long-wavelength modulation of the rotationally misaligned top graphene sheet on the surface of highly-oriented pyrolytic graphite. Nearly two decades later, the first spectroscopic study of these naturally occurring \moire materials systematically correlated the twist angle from STM topography with spectroscopic information about the system's electronic structure \cite{li2010observation}.

\hspace{2mm} This early spectroscopic study was followed by the seminal work of Rafi Bistritzer and Allan MacDonald \cite{bistritzer2011moire}, predicting the emergence of flat \moire bands at an infinite sequence of ``magic angles'' in twisted bilayer graphene, which contradicted the contemporary expectation that the bandwidth of \moire bands would decrease monotonically with twist angle. Instead, a delicate balance of the intralayer hopping terms and the out-of-plane tunneling terms in the Hamiltonian of twisted bilayer graphene produces flat bands at the magic angle of 1.1$^\circ$, away from the zero twist angle limit. As the angle is further reduced from this magic angle, the bandwidth of these lowest energy bands increases.

\hspace{2mm} The ``Bistritzer-MacDonald (BM) continuum model’’ produced a quantitatively accurate description of the electronic band structure of twisted bilayer graphene at small, incommensurate twist angles. Shortly thereafter, STM studies found van Hove singularities in twisted bilayer graphene made from polycrystalline graphene grown by chemical vapor deposition \cite{wong2015local}. By comparing the local density of states to the integrated density of states of the BM model, this study demonstrated quantitative consistency across a range of twist angles, marking the first experimental verification of the BM model that underpins nearly all theoretical \moire material models today.

\hspace{2mm} Within this new era of \moire materials research, new micrometer- and nanometer-scale angle-resolved photoemission spectroscopy (microARPES, nanoARPES) studies have produced direct observations of the generically flat \moire bands in both twisted bilayer graphene (Fig. \ref{fig:FlatBands}a), twisted trilayer graphene, and twisted transition metal dichalcogenides \cite{utama2021visualization,lisi2021observation,chen2023strong,li2022observation,stansbury2021visualizing}. These studies show narrow electronic bands (tens of meV in bandwidth) that are localized in momentum near each valley. By directly visualizing the band dispersions of \moire materials, these experiments further support our understanding of the quantitative accuracy of continuum models in capturing the electronic structure of \moire materials, despite their simplifying approximations.


\subsection{Cascades of Electronic Transitions}

\hspace{2mm} Correlation effects in \moire materials are dominant, and often cause strong reorganizations of the material's low-energy excitations. In MATBG, these effects are observed even at relatively high temperatures (T $> 6$ K), above the transition temperatures of the system's superconducting and magnetic topological phases. At these high temperatures, local spectroscopic and compressibility probes established the fundamental properties of a correlated metallic regime in MATBG, akin to a parent phase in this material, which sets the stage for various low-temperature ground states \cite{wong2020cascade,zondiner2020cascade}.

\hspace{2mm} Fig. \ref{fig:FlatBands}b shows STS data of the differential conductance \dIdVSG in MATBG at T = $6$ K. When the chemical potential lies above (\filling \: $> +4$) or below (\filling \: $< -4$) the flat bands in energy, STS measurements resolve the van Hove singularities (vHs) of the two flat bands of MATBG as two sharp, parallel lines in \dIdVSG (top and bottom regions; Fig. \ref{fig:FlatBands}b), as observed by many groups \cite{xie2019spectroscopic, kerelsky2019maximized, choi2019electronic, jiang2019charge}. When the chemical potential lies within one of the flat bands ($-4 < $ \filling \: $< +4$; Fig. \ref{fig:FlatBands}a), STS spectra show immediate broadening of the flat bands, a consequence of strong electronic interactions \cite{xie2019spectroscopic}, further accompanied by a sequence of finer features (the so-called ``cascade'' features) observed near \EF \cite{wong2020cascade}. Cascade features repeat in evenly spaced filling intervals, where peaks first appear near \EF and emanate in energy as a function of density between integer fillings. Similar features were observed in magic-angle twisted trilayer graphene \cite{kim2022evidence}.

\hspace{2mm} The cascade of transitions observed in STS indicates a reorganization of the low-energy excitations of MATBG throughout the entire bandwidth and density range of the flat bands. When the flat bands are partially filled, the energies of the system's excitations (as probed by STS in this strongly correlated regime) depend heavily on the number of carriers in the system. The cascade features can be understood as excitations on the correlated metallic phase that forms the parent phase of MATBG \cite{kang2021cascades,song2021matbg,datta2023heavy}. Proposals of an unusual dichotomy of density-dependent heavy and light quasiparticle excitations, the so-called heavy fermion picture of MATBG, are derived from these spectroscopic observations, and do not require isospin symmetry-breaking effects to explain the cascade features. In this picture, localized flat-band ``f'' states hybridize with itinerant topological ``c'' states in a simplified and predictive model of MATBG that includes all of the relevant symmetries of the system \cite{song2021matbg}. Some of the finer, density-dependent sharp peak features of spectroscopy have been attributed to a Kondo-like effect that would reflect the interaction between light delocalized and heavy localized quasiparticles \cite{hu2023symmetric, zhou2023kondo, huang2023evolution, chou2023kondo}. More experimental measurements are required to fully establish this heavy fermion picture, which contrasts a symmetry-breaking picture of the parent phase that proposes a sequence of long-range isospin flavor-polarized metallic phases separated by first or second order phase transitions, as driven by Stoner-type electronic instabilities \cite{zondiner2020cascade}.

\hspace{2mm} Similarly, scanning SET measurements of the compressibility of MATBG (Fig. \ref{fig:FlatBands}c) show a repetitious sequence of sawtooth features near integer fillings, which are particularly prominent in the upper flat band (\filling \: $> 0$) \cite{zondiner2020cascade}. Sawtooth features occur at transitions between highly compressible, large density of states regions of the many-body spectrum to highly incompressible, small density of states regions. Oscillations in the chemical potential are also observed in STS measurements of the higher energy ``remote'' band features away from \EF \cite{wong2020cascade}, although STS measurements are less direct than those of scanning SET.

\hspace{2mm} The sudden drops in compressibility when increasing (decreasing) the carrier density of MATBG across positive (negative) integer fillings parallel the Landau level asymmetry reported in magneto-transport measurements \cite{cao2018unconventional, yankowitz2019tuning, lu2019superconductors, saito2020independent, stepanov2020untying}. There, Landau fan features from non-zero integer fillings emanate in only one direction, pointing away from charge neutrality. Chemical potential measurements offer a natural interpretation of these unusual features as derived from the asymmetric dispersion of either the system's reorganized isospin subbands \cite{tarnopolsky2019origin, bultinck2020ground} or of its strong-coupling excitation spectrum \cite{kang2021cascades,song2021matbg,datta2023heavy}. Under both scenarios, quantum oscillations are only resolvable in the new low-density-of-states / low compressibility regions of the spectrum, while none are observable in the high-density-of-states / highly compressible regions. The persistence of such a reorganization of electronic states to high temperatures (T $>$ $16$ K; Fig. 1c) makes symmetry-breaking dominated mechanisms less likely, but are perhaps better understood as a consequence of the formation of a strongly correlated state, such as that captured by the heavy fermion picture of MATBG \cite{kang2021cascades,song2021matbg,datta2023heavy}.

\hspace{2mm} The combination of early magneto-transport experiment results with the observations presented in this section \cite{wong2020cascade, zondiner2020cascade} have helped identify the most fundamental electronic ingredients in MATBG, which provide crucial constraints for developing microscopic theories derived from a holistic experimental understanding of this correlated system \cite{kang2021cascades,song2021matbg}.

\section{\label{sec:Corr_Graphene}Correlated Insulating Phases in \Moire Graphene}
\subsection{Isospin Symmetry-Breaking in \Moire Graphene}

\setlength{\columnsep}{0pt}
\setlength{\intextsep}{0pt}
\begin{figure*}
    \centering
    \includegraphics[width=\textwidth]{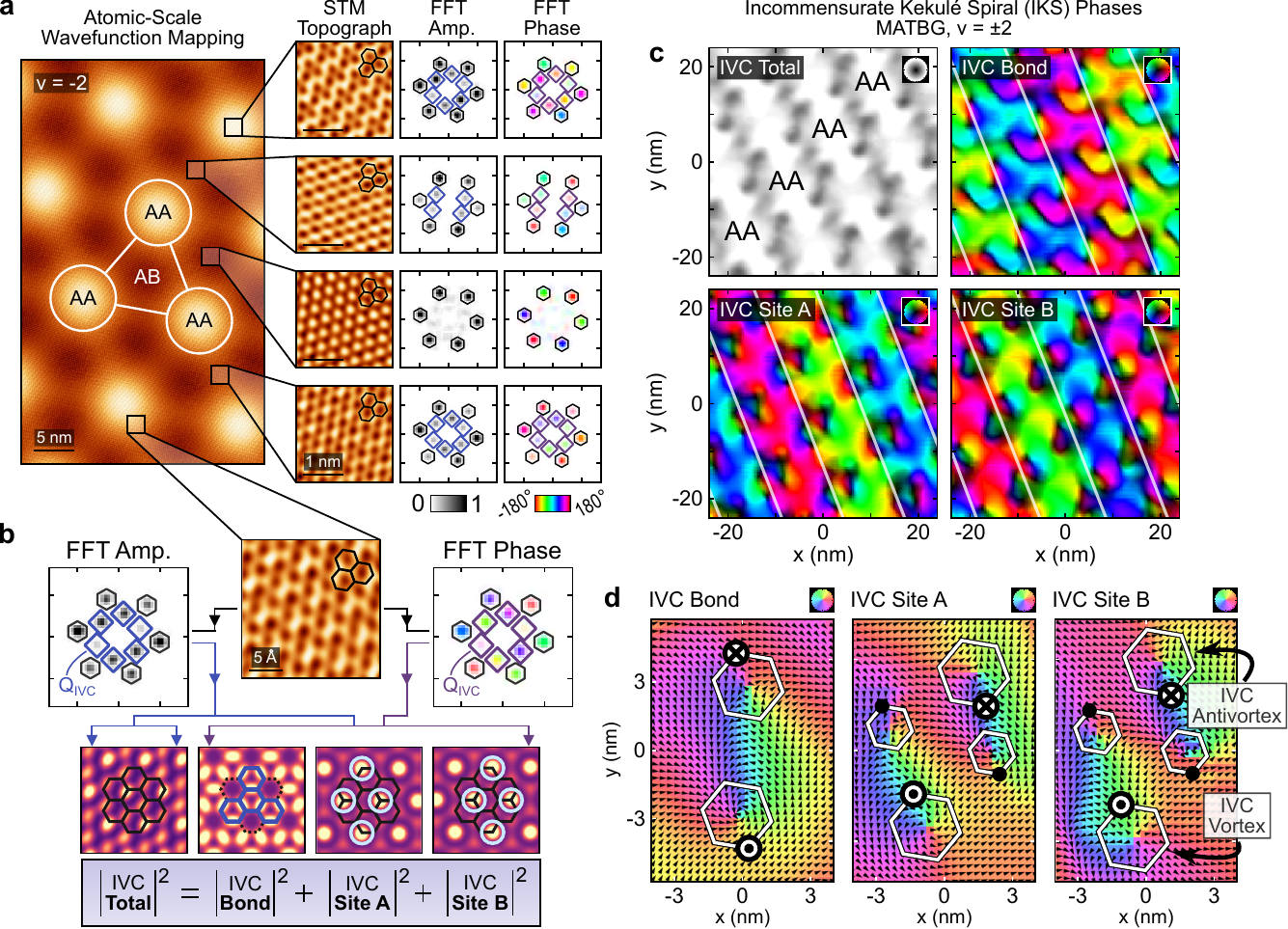}
    \caption{\label{fig:Corr_Graphene} \textbf{Mapping Correlated Insulators in \Moire Graphene.} \textbf{a,} Large-scale ($50$ x $30$ nm$^2$) atomically resolved STM topographic image obtained at filling \filling \: = $-2$ in MATBG, which shows atomic-scale signatures of intervalley coherence (IVC) that modulate on the moir{\'e}-scale. Small-scale ($2$ x $2$ nm$^2$) zoomed-in images show how these wavefunction patterns and their FFTs change among locations. \textbf{b,} Schematic diagram of the local FFT order parameter decomposition of STM images. Small-scale ($0.25$ - $1$ nm$^2$) topographic images, cropped from the large-scale image in \textbf{a}, are used to extract local FFT amplitude (left) and phase (right) information. This information is further decomposed into three IVC order parameters, whose magnitudes sum in quadrature to produce the IVC total order parameter (bottom row). \textbf{c,} Local order parameter maps of IVC total (upper left panel), which describes the overall intensity of the IVC wavefunction patterns, and of IVC Bond (upper right) / Site A (lower left) / Site B (lower right), the magnitude of which characterizes the type of IVC wavefunction patterns and the phase of which characterizes the shape of the patterns. \textbf{d,} Local order parameter maps of IVC vortices and antivortices in MATBG, where the complex phase of the local IVC order parameters wind by $\pm 2 \pi$. Reprinted from \cite{nuckolls2023quantum} with minor adaptations for style and formatting purposes.
    }
\end{figure*}

\hspace{2mm} Spontaneous symmetry-breaking occurs when Coulomb interactions among electrons set the dominant energy-scale in the system's Hamiltonian, producing symmetry-broken phases that mitigate these interaction energetic costs. In \moire materials, this motif describes an abundance of correlation-driven insulators observed at commensurate integer fillings in magic-angle twisted bilayer graphene (MATBG; Fig. \ref{fig:PhaseDiagrams}b) \cite{cao2018correlated, yankowitz2019tuning, lu2019superconductors}, magic-angle twisted trilayer graphene (MATTG) \cite{park2021tunable, hao2021electric}, twisted monolayer bilayer graphene (tMBG; Fig. \ref{fig:PhaseDiagrams}c) \cite{polshyn2020electrical, chen2021electrically}, and twisted double bilayer graphene (tDBG) \cite{liu2020tunable, cao2020tunable, shen2020correlated}.

\hspace{2mm} In this section, we highlight work that has precisely identified complex symmetry-broken orders in MATBG and MATTG using the spatial- and energy-resolution of scanning tunneling microscopy / spectroscopy (STM / STS) \cite{nuckolls2023quantum,kim2023imaging}. In MATBG and MATTG, correlated insulators present signatures of intricate symmetry-broken orders imprinted upon the local density of states on the graphene atomic-scale. High-resolution STM imaging experiments were particularly important for understanding the role of isospin symmetry-breaking effects in these platforms in the absence of any predicted differentiating signatures accessible to globally averaged probes.

\hspace{2mm} Prior to these recent STM studies, transport experiments on MATBG first revealed a correlated insulator at \filling \: $= -2$, half-filling of the system's lower flat band \cite{cao2018correlated}. These observations were suggestive of a Mott insulator, particularly when observed in proximity to low-density superconducting phases at nearby densities (i.e. reminiscent of the cuprate phase diagram) \cite{shirane1987two, chakravarty1988low, manousakis1991spin}. Similar observations have been made in MATTG near \filling \: $= -2$ \cite{park2021tunable, hao2021electric}. Further transport experiments, however, provided contrasting insights into the nature of these insulators. First, subsequent work uncovered correlated insulators at odd integer fillings (ex. \filling \: $\pm 3$ in MATBG) \cite{yankowitz2019tuning, lu2019superconductors, saito2020independent}, emphasizing the importance of valley degrees of freedom in MATBG that, together with the spin degree of freedom, supports a $4$-fold isospin degeneracy. Second, subsequent work confirmed observations of a $2$-fold reduced-degeneracy Landau fan observed in magneto-transport measurements emanating from \filling \: $= -2$ \cite{yankowitz2019tuning, lu2019superconductors, saito2021hofstadter}, in contrast to the $4$-fold degenerate Landau fan observed near charge neutrality. Consistently observed across samples, this was a strong indicator of a broken-isospin-symmetry insulator at \filling \: $= -2$ \cite{tarnopolsky2019origin, bultinck2020ground, kang2019strong}, distinct from the originally proposed Mott phase.


\subsection{Mapping Correlated Insulators in \Moire Graphene}

\hspace{2mm} One outstanding question was that none of these experiments yields information about \textit{which} isospin symmetries are broken near \filling \: $= \pm 2$. Answering this question is important for addressing the nature of superconductivity in MATBG and MATTG, as superconducting phases appear most robustly near \filling \: $= \pm 2$. As is often the case, the origins of many-body electronic phases are elusive to global probes, where the only distinguishable signatures of symmetry-breaking are imprinted upon the state’s microscopic wavefunction. To address this issue, the spatial resolution of the STM provided an incisive probe of the exact nature of symmetry-breaking effects in MATBG and MATTG by characterizing atomic-scale signatures of their electronic phases \cite{nuckolls2023quantum, kim2023imaging}. Such techniques have not only enabled us to narrow the number of candidate ground states consistent with experiments by visualizing the symmetries being broken, but also to identify the correlated insulators at \filling \: $= \pm 2$ in these two platforms by closely comparing local observables to those of theoretical ground state candidates.

\hspace{2mm} Shown in Fig. \ref{fig:Corr_Graphene}a is a low-bias STM image of the \filling \: $= -2$ correlated insulator in MATBG, with four atomic-scale images of high-symmetry stacking regions of the \moire superlattice. Overlaying the graphene atomic lattice is an atomic-scale wavefunction pattern (i.e. $|\psi(x)|^2$) dubbed the ``R3 pattern'', which breaks translation symmetry, forming a $\sqrt{3}$ x $\sqrt{3}$ reconstruction that triples the graphene unit cell, and breaks rotational symmetries, forming locally nematic R3 patterns even in high-symmetry sites of the \moire superlattice where one expects the rotational symmetries of graphene to be preserved \cite{nuckolls2023quantum}. Similar patterns appear in MATTG near \filling \: $= -2$ \cite{kim2023imaging}, and within the zeroth Landau level (\filling \: $= 0$) of monolayer graphene \cite{liu2022visualizing}.

\hspace{2mm} The R3 pattern is most intuitively understood in Fourier space (local magnitude and phase fast Fourier transform (FFT) plots, Fig. \ref{fig:Corr_Graphene}a). In each local FFT, the six outer peaks are the Bragg peaks of graphene ($Q_{Bragg}$), associated with the wavevector $\vec{G}$. The six inner peaks ($Q_{IVC}$) appear near \filling \: $= \pm 2$, and are associated with a longer wavelength modulation than the Bragg peaks. They identify the $\sqrt{3}$ x $\sqrt{3}$ reconstruction of the R3 pattern, and are a smoking gun signature of intervalley coherent (IVC) isospin order because they are associated with the wavevector $\vec{K}$, which points from valley $\vec{K}$ to valley $\vec{K'}$. Note that STM measurements of $|\psi(x)|^2$ of the R3 pattern reflect relative phase information between valleys contained in the wavefunction $\psi(x)$.

\hspace{2mm} In MATBG, a symmetry-based, group-theoretical approach was introduced to extract locally defined order parameters from STM images, which enabled classifying the symmetries of its correlated phases. These local order parameters (Fig. \ref{fig:Corr_Graphene}b) are complex-valued functions that capture all of the translation and rotational transformation properties of the R3 pattern under the symmetries of the graphene lattice. Particularly, the IVC order parameters measure the intensity of specific types and shapes of symmetry-broken R3 patterns. IVC Bond / Site A / Site B order parameters (Fig. \ref{fig:Corr_Graphene}b) measure the intensity of the R3 pattern with states centered on the carbon-carbon bonds / A-sublattice carbon site / B-sublattice carbon site of graphene. The magnitude of each complex order parameter measures the intensity of that type of R3 pattern, while the complex phase measures the shape of the R3 pattern of a given type within the newly tripled graphene unit cell.

\hspace{2mm} Over a large field-of-view STM image of the \filling \: $= -2$ correlated insulator, Fig. \ref{fig:Corr_Graphene}c shows a gray-scale plot of the real-valued IVC Total order parameter (upper left panel) and color plots of the complex-valued IVC Bond (upper right), IVC Site A (lower left), and IVC Site B (lower right) order parameters. The color of each pixel represents the phase of these order parameters while the brightness represents their magnitude at each location. The IVC Total map, which measures the overall strength of the R3 pattern, is moir{\'e}-periodic; whereas, the IVC-Bond, IVC-A, and IVC-B maps break \moire translation and rotational symmetries, showing stripe-like features that change in phase as a function of position in the \moire superlattice. Moreover, IVC order parameter vortices and antivortices (phase winding $\Delta \theta = \pm 2 \pi$ over a closed loop) appear upon subtracting a linear phase background from IVC Bond / A / B maps (Fig. \ref{fig:Corr_Graphene}d). These vortex features appear near locations of deep suppressions of the IVC Total strength, located near AB- / BA-stacking regions of the \moire superlattice. Through careful comparison of these local observables with simulations of candidate ground states in MATBG, STM imaging experiments identified that incommensurate Kekul{\'e} spiral (IKS) states \cite{kwan2021kekule, wagner2022global, wang2022kekul,hong2022detecting}, highly complex quantum phases that exhibit multi-scale symmetry-breaking, are the most likely candidate ground states at \filling \: $= \pm 2$ in MATBG in typical strained devices ($\epsilon$ $>$ 0.1$\%$) \cite{nuckolls2023quantum}. With a separate approach, high-resolution STM imaging experiments on MATTG have detected a subtle mismatch of the measured and ideal $Q_{IVC}$ wavevectors \cite{kim2023imaging}. These measurements are also indicative of moir{\'e}-scale translation symmetry-breaking, and are suggestive of IKS order in MATTG. This interpretation was further corroborated by unusual modulations in the atomic-scale wavefunction patterns of MATTG, as measured in an auto-correlation analysis that uncovered differences in the R3 pattern between neighboring \moire unit cells.

\hspace{2mm} Finally, the R3 patterns observed in both MATBG and MATTG were found to be particularly sensitive to external tuning parameters. For example, the R3 pattern in MATBG appears sensitive to the presence of interlayer heterostrain, where a few ultra-low strain ($\epsilon$ $<$ 0.1$\%$) devices showed R3 patterns that do not exhibit IVC stripe and vortex features, instead presenting moir{\'e}-periodic patterns that locally resembles a distinct, time-reversal-invariant IVC (TIVC) order \cite{kang2019strong, caluguaru2022spectroscopy}. A full understanding of these observations and the ultra-low strain limit of MATBG is an active direction of research, both experimentally and theoretically \cite{kwan2023electron,christos2023nodal}. Moreover, the R3 patterns in MATBG and MATTG appear sensitive to carrier density. While the superconducting and pseudogap phases in both MATBG and MATTG show R3 patterns indicative of IVC order, their R3 patterns are distinguishable from those of the correlated insulator at \filling \: $= \pm 2$. In MATBG, discontinuous reorganizations of the R3 pattern were observed as the correlated insulating gap near \filling \: $= \pm 2$ closed and reopened into the proximate superconducting or pseudogap states, both showing new IVC order that remains unidentified to-date \cite{nuckolls2023quantum}. In MATTG, the moir{\'e}-scale wavevectors of the R3 pattern changes continuously with doping, with no discontinuity observed between the insulator and proximate superconducting / pseudogap phase \cite{kim2023imaging}.

\hspace{2mm} In total, these two works represent an unprecedentedly precise visualization of the many-body wavefunctions in correlated electron systems. They provide illustrative examples of the incisive capabilities of wavefunction mapping with the STM to establish complex correlated quantum states. Such capabilities will likely be useful in decisively identifying the microscopic nature of correlated phases not only in other graphene-based \moire materials, but also those found in semiconducting transition metal dichalcogenide \moire materials.

\section{\label{sec:Corr_TMD}Correlated Insulating Phases in \Moire TMDs}
\subsection{Electron Crystal Phases}

\setlength{\columnsep}{0pt}
\setlength{\intextsep}{0pt}
\begin{figure*}
    \centering
    \includegraphics[width=\textwidth]{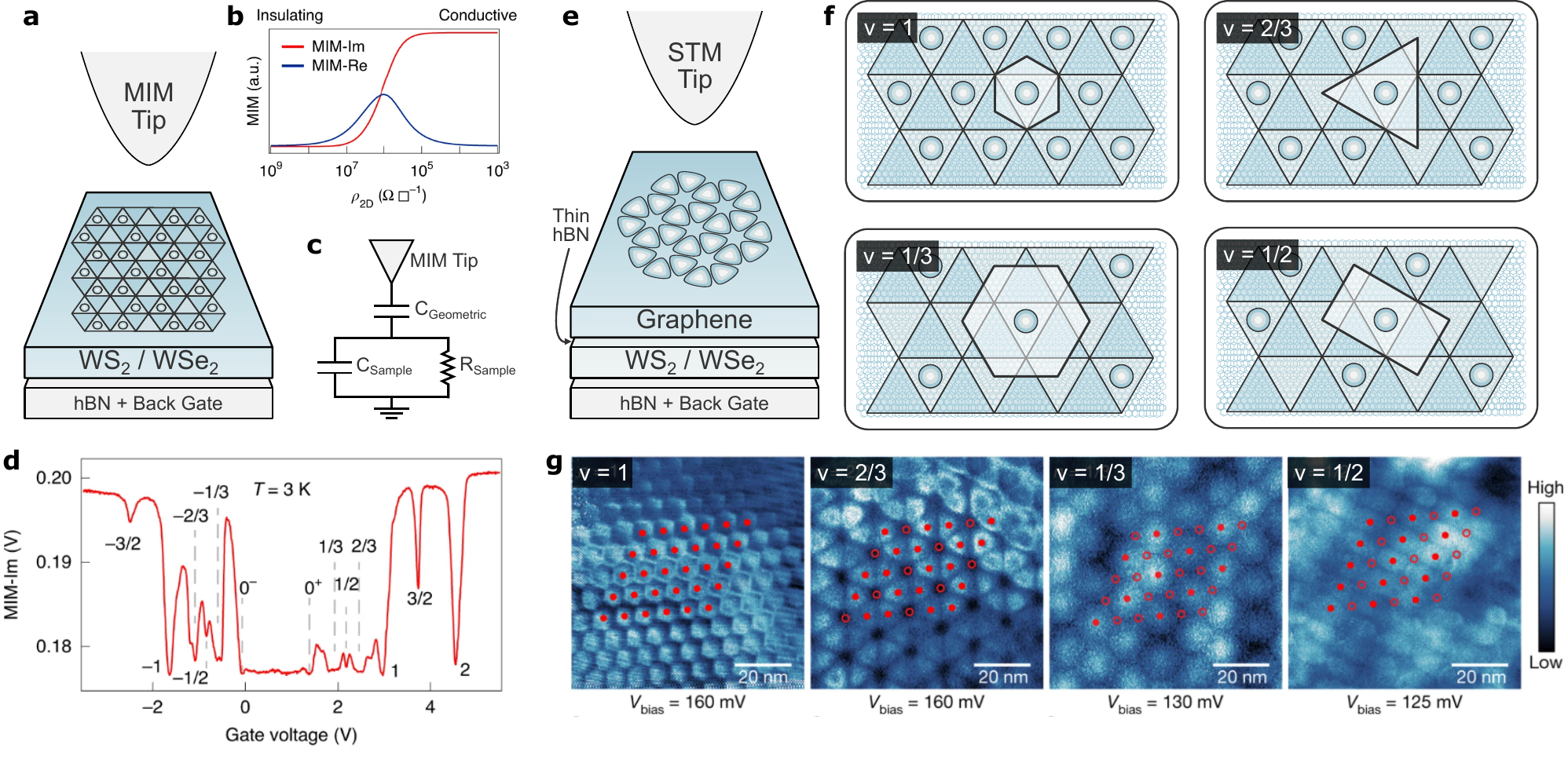}
    \caption{\label{fig:Corr_TMD} \textbf{Imaging Correlated Insulators in \Moire TMDs.} \textbf{a,} Schematic diagram of a microwave impedance microscopy (MIM) experiment on aligned \WSe /  WS$_2$. \textbf{b,} Typical response curve in MIM experiments. A microwave AC signal impinges upon the sample through a sharp metallic tip, and the complex tip-sample impedance is measured in the reflected signal. The imaginary part of the MIM signal (MIM-Im) is related to the local conductivity of the sample. \textbf{c,} Effective circuit model for the tip-sample impedance measured in \textbf{b}. \textbf{d,} MIM-Im signal measured as a function of gate voltage. Deep suppressions of MIM-Im are indicative of resistive dissipation of the signal within the sample, which occur at Mott and generalized Wigner crystal (GWC) states in \WSe /  WS$_2$. \textbf{e,} Schematic diagram of an STM charge-sensing experiment. \WSe / \WS is covered by thin hBN and a graphene sensing layer. \textbf{f,} Schematic diagrams of the charge configuration of the Mott insulator at \filling \: = $1$ (upper left panel) and GWC states at \filling \: = $2/3$, $1/3$, and $1/2$ (right and lower panels). Dark outlined regions indicate the \moire superlattice area per electron (not necessarily the state’s unit cell), as measured in STM charge-sensing experiments. For fillings \filling \: = $1$, $1/3$, and $1/2$, this outlined region is equivalent to the enlarged Wigner-Seitz \moire unit cell of the correlated state. \textbf{g,} Charge-sensing dI/dV maps of the encapsulating graphene sensing layer. The spatial characteristics of local discharging events in graphene directly reflect the charge configuration of correlated state in the underlying \WSe / WS$_2$. Reprinted from \cite{huang2021correlated,li2021imaging} with minor adaptations for style and formatting purposes.
    }
\end{figure*}

\hspace{2mm} An effective way to lower the free energy of a strongly interacting system, particularly at very low densities, is for electrons to order themselves into crystalline solid phases called ``Wigner crystals'', first proposed by Eugene Wigner in the 1930's \cite{wigner1934interaction,monarkha2012two}. The Coulomb repulsive interaction between electrons is long-ranged, and diverges in the limit where electronic wavefunctions overlap. The physical separation of electrons in a well-organized lattice mitigates this short-range divergence. Schematically, the average Coulomb potential energy between electrons scales inversely with the average distance $a$ between electrons ($U \propto 1/a$); whereas, the kinetic energy of each electron scales inversely with the average inter-particle distance squared ($K \propto \hbar^2 / m a^2$). Thus, in the low-density limit, interactions promote the energetic significance of the system's local geometric arrangement of electrons. Evidence for classical Wigner crystals were first reported nearly fifty years after their original proposal in dilute electron gases on the surface of liquid helium \cite{grimes1979evidence}. Evidence for quantum Wigner crystals were reported at small Landau level filling factors of GaAs / AlGaAs quantum wells \cite{andrei1988observation}. These and many other studies have provided evidence for Wigner crystals using globally averaged techniques, which makes it hard to distinguish such states from disorder-pinned localized phases. More recently, local probe techniques using the STM have directly imaged Wigner crystals in the lowest Landau level of bilayer graphene, uncovering their hexagonal close-pack structures and probing their thermal and quantum melting transitions \cite{tsui2023direct}.


\subsection{Charge-Ordered Phases in \Moire Transition Metal Dichalcogenides}

\hspace{2mm} Recently, the flat electronic bands of \moire materials have produced new opportunities to realize a variety of electron crystal phases at zero magnetic field \cite{regan2020mott}. These so-called ``generalized Wigner crystal’’ (GWC) phases differ from those of the fractional quantum Hall regime because of their interaction with the \moire superlattice potential. Specifically, GWCs break discrete rather than continuous translational symmetry, with spatial structures dominated by the \moire superlattice.  In semiconducting transition metal dichalcogenide (TMD) \moire materials, spectroscopic studies have measured deep \moire electrostatic potentials \cite{shabani2021deep}, an ideal template for localizing electrons within the superlattice at specific, rational filling factors of electrons per \moire unit cell. At rational fillings, a natural partial occupancy of \moire sites is possible that, for example, triples the superlattice unit cell at one-third filling (Fig. \ref{fig:PhaseDiagrams}e) \cite{xu2020correlated,huang2021correlated}. At fillings \filling \: $= \pm 1/3$, $\pm 1/4$, and $\pm 1/7$, isotropic triangular GWCs appear on the triangular \moire superlattice with enlarged and rotated primitive lattice vectors that break moir{\'e}-scale translation symmetry. At fillings \filling \: $= \pm 1/2$, $\pm 2/5$, GWCs exhibit enlarged unit cells that additionally break the rotational symmetries of the superlattice (i.e. moir{\'e}-scale ``stripe phases’’) \cite{jin2021stripe}. Experimentally, filling-dependent resistive features should identify these phases in transport measurements. However, the semiconducting bandgap of monolayer TMDs causes a Schottky barrier at the interface between these materials and their metallic device contacts, inhibiting direct measurements. Consequently, identifying correlated phases in TMD \moire materials using low-temperature transport techniques has been notoriously difficult.

\hspace{2mm} In this Section, we highlight recent local probe experiments using microwave impedance microscopy (MIM) and scanning tunneling microscopy (STM), which probe charge-ordered phases in \WSe / \WS \moire superlattices while circumventing the complications of ohmic contact engineering. These experiments uncover an abundance of GWCs at nearly two dozen fractional commensurate fillings of the \moire superlattice, and even provide spatial signatures of these phases that directly confirm their translation and rotational symmetry-broken nature.


\subsection{Imaging Correlated Insulators in \Moire TMDs}

\hspace{2mm} The earliest reports of charge-ordered states in TMD-based \moire materials investigated \WSe / \WS \moire superlattices, primarily using optical probes \cite{tang2020simulation,regan2020mott}. These reports uncovered a correlated insulator at filling \filling \: $= -1$, whose integer charge occupancy per \moire unit cell was suggestive of a Mott insulator (upper left panel; Fig. \ref{fig:Corr_TMD}f). This was further evidenced by signatures of local, antiferromagnetically ordered moments at \filling \: $= -1$, as expected in the triangular lattice Hubbard model \cite{tang2020simulation}. Additionally, correlated insulators were observed near fractional fillings \filling \: $= -1/3$ and $-2/3$ \cite{regan2020mott}. The simple fractional values of charge occupancy per \moire unit cell of these insulators was suggestive of GWCs (upper right and lower left panels; Fig. \ref{fig:Corr_TMD}f).

\hspace{2mm} Following these initial reports, a study using MIM \cite{huang2021correlated}, concurrent with a newly established ``exciton sensing'' optical study \cite{xu2020correlated}, reported an abundance of GWCs occurring at roughly 20 rational filling factors (Fig. \ref{fig:PhaseDiagrams}e). In MIM experiments, a sharp metallic tip capacitively couples to the sample (Fig. \ref{fig:Corr_TMD}c). A microwave signal (typically in the range of $1$ to $10$ GHz) impinges upon the sample, and the imaginary part of the reflected signal (denoted MIM-Im) provides information about the signal's resistive dissipation within the sample (Fig. \ref{fig:Corr_TMD}b) \cite{cui2016quartz}. Here, the MIM-Im signal strength acts effectively as a local transport probe of the GWC states. Because the microwave-induced AC electric field is local, only the carriers nearest the tip-sample junction are affected. Thus, the sample contact is only necessary for electrostatic gating and, therefore, is amenable to the large RC charging time constants typical of non-ohmic contacts.

\hspace{2mm} Fig. \ref{fig:Corr_TMD}d shows the MIM-Im signal of this experiment as a function of back gate voltage. Deep suppressions of the MIM-Im signal act as signatures of correlated insulators that appear localized to simple integer and rational filling factors of the flat band of \WSe / WS$_2$ \cite{huang2021correlated}. These deep suppressions appeared at nearly the same gate voltages at several locations in their device. In addition, temperature-dependent studies of this signature demonstrated that the insulating suppressions vanished at filling-dependent temperatures that resembled those expected from small-scale Monte Carlo simulations of the proposed GWC states.
    
\hspace{2mm} Subsequently, STM imaging experiments provided unequivocal evidence for these Mott and GWC states by introducing a new charge-sensing method (Fig. \ref{fig:Corr_TMD}e) that enabled researchers to directly visualize these correlated ground states in real-space \cite{li2021imaging}. In these experiments, aligned \WSe / \WS was encapsulated from the top by a thin hBN spacer layer, covered by a monolayer graphene sensing layer that was exposed to the STM tip. When \WSe / \WS is electrostatically gated into a correlated insulating state, discrete charges become localized to high-symmetry-stacking regions of the \moire superlattice (Fig. \ref{fig:Corr_TMD}f). Each of these trapped charges can be locally discharged by a voltage-biased STM tip, and this discharging event can be measured in dI/dV spectra of the top graphene sensing layer. Thus, this technique provides a direct measurement of the \moire superlattice area per trapped electron / hole (not necessarily the state's Wigner-Seitz unit cell), as schematically depicted in the panels of Fig. \ref{fig:Corr_TMD}f by dark outlined polygons.

\hspace{2mm} As observed in STM charge-sensing experiments, states at fillings \filling \: $= 1/3$, $2/3$ (middle panels; Fig. \ref{fig:Corr_TMD}g) are indeed triangular and hexagonal GWCs, respectively, that break \moire translation symmetry. In addition, the GWC at filling \filling \: $= 1/2$ (right panel; Fig. \ref{fig:Corr_TMD}g) is a moir{\'e}-scale stripe phase that breaks both \moire translation and rotational symmetries.

\section{\label{sec:Topo}Topological Insulating Phases}
\subsection{Time-Reversal Symmetry in \Moire Materials}

\setlength{\columnsep}{0pt}
\setlength{\intextsep}{0pt}
\begin{figure*}
    \centering
    \includegraphics[width=\textwidth]{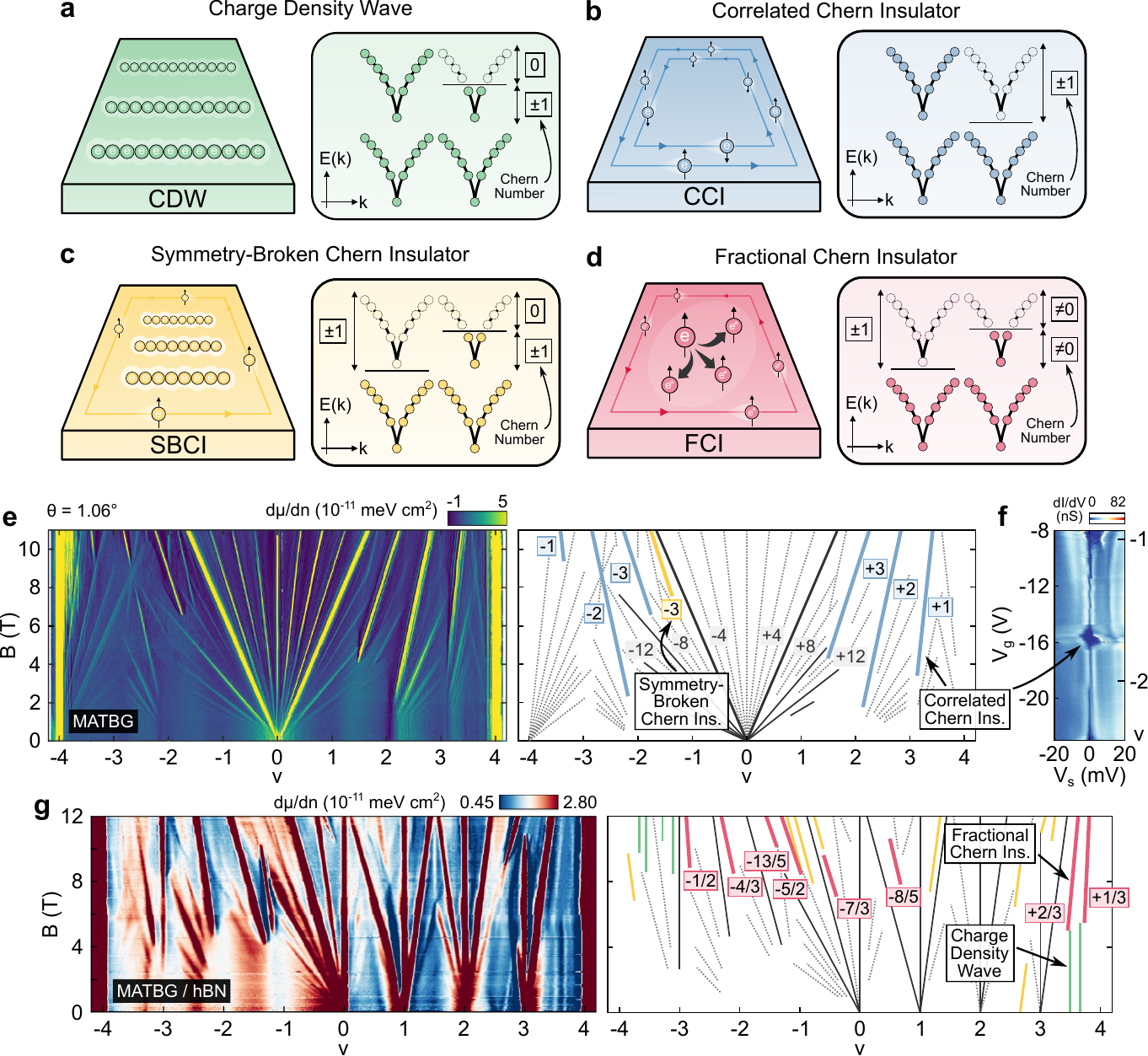}
    \caption{\label{fig:Topo} \textbf{Local-Sensing of Correlation-Driven Topological Phases in \Moire Graphene.} \textbf{a,} Schematic of a charge density wave (CDW) state, which breaks translation symmetry, but preserves time-reversal-symmetry (T-symmetry), which protects the Dirac points of MATBG. Thus, the CDW state is topologically trivial. \textbf{b,} Schematic of a correlated Chern insulator (CCI), which breaks T-symmetry, and is thus topologically non-trivial. \textbf{c,} Schematic of a symmetry-broken Chern insulator (SBCI), a topological phase that breaks both T-symmetry and translation symmetry. \textbf{d,} Schematic of a fractional Chern insulator (FCI), a topological phase that breaks T-symmetry and hosts non-trivial topological order. \textbf{e,} Inverse compressibility d$\mu$/dn data of MATBG measured using scanning SET. CCIs (purple lines) and a SBCI (yellow line) are observed as sharp incompressible peaks in d$\mu$/dn, which shift in density as a function of magnetic field according to their respective Chern numbers. \textbf{f,} Differential conductance dI/dV of MATBG measured using STS. A large spectroscopic gap indicative of a CCI with \Chern \: $= -3$ (as measured in STS by its magnetic-field dependence) appears in the middle of a \moire flat band, unassociated with single-particle dI/dV features. \textbf{g,} Same as \textbf{e}, measured on MATBG aligned to hBN. CDWs (green lines), SBCIs (yellow lines), and FCIs (red lines) are observed as sharp incompressible peaks in d$\mu$/dn, which shift in density as a function of magnetic field according to their respective Chern numbers. Reprinted from \cite{yu2022correlated, nuckolls2020strongly,xie2021fractional} with minor adaptations for style and formatting purposes.
    }
\end{figure*}

\hspace{2mm} Magnetic topological phases arise in \moire materials in ways different than in the quantum Hall regime \cite{halperin2020fractional, tong2016lectures} because the origins of the flat bands in these two systems are fundamentally different. \Moire flat bands arise from either the hybridization or Brillouin zone folding of the low-energy dispersive bands of the system's 2D layers. Such effects produce narrow electronic bands at zero magnetic field \cite{morell2010flat, bistritzer2011moire} with intrinsic Berry curvature derived from the material's component layers. These flat \moire bands preserve time-reversal symmetry, in contrast to the magnetic-field derived flat Landau levels in quantum Hall systems. Breaking time-reversal symmetry in \moire materials can unlock this hidden Berry curvature, stabilizing magnetic topological states even in the near absence of $d$-electron states or spin-orbit coupling (ex. graphene-based \moire materials).

\hspace{2mm} In this Section, we highlight the discoveries of new correlation-driven magnetic topological phases \cite{rachel2018interacting}, many of which were first discovered using scanning SET and STM / STS \cite{nuckolls2020strongly, choi2021correlation, yu2022correlated, pierce2021unconventional, xie2021fractional, kometter2022hofstadter, foutty2023mapping}. In the subsequent Section (\ref{sec:Ferro}), we will discuss local imaging experiments of magnetic topological phases using scanning nano-SQUID.


\subsection{Orbital Magnetism in Graphene-Based \Moire Materials}

\hspace{2mm} In graphene-based \moire materials, topology stems from the low-energy dispersion of monolayer graphene, characterized by linear crossings called ``Dirac points'' that host singularities of Berry curvature localized at the $\vec{K}$ and $\vec{K'}$ points of graphene's Brillouin zone. These Dirac points are protected by a composite $C_2T$-symmetry, comprised of an $180^\circ$ rotation ($C_2$) and a spinless version of time-reversal symmetry ($T$). In MATBG, breaking either symmetry gaps the Dirac points, producing an emergent set of flat valley-Chern subbands. Asymmetrically populating a subset of these Chern bands with a net Chern number then gives rise to an anomalous Hall effect in MATBG that can be measured directly in transport measurements of $\rho_{xy}$.

\hspace{2mm} The anomalous Hall effect was first observed near \filling \: $= +3$ in MATBG devices aligned to hBN (MATBG / hBN), which breaks the $C_2$-rotational-symmetry of graphene \cite{sharpe2019emergent}. MATBG / hBN showed an enormous, hysteretic anomalous Hall effect below $5$ K and a drop in the longitudinal resistance ($R_{xx}$), signifying the onset of an incipient Chern insulator, and non-local transport consistent with the presence of a chiral edge mode expected of an orbital magnetic state. Subsequent work showed that in a more homogeneous MATBG / hBN device (both in twist angle and in charge homogeneity), a deep suppression of the longitudinal resistance near \filling \: $= +3$ appeared concomitant with a quantized Hall resistance of $\frac{h}{e^2}$, indicative of a quantum anomalous Hall insulator at this filling. 

\hspace{2mm} In addition to these reports in MATBG, contemporaneous work has demonstrated anomalous Hall responses \cite{chen2020tunable, polshyn2020electrical, chen2021electrically} or strong non-linear Hall responses \cite{stepanov2020untying, liu2020tunable} in several other \moire materials. Particularly noteworthy is the Chern insulating phase at \filling \: $= -1$ in ABC-stacked trilayer graphene aligned to hBN (\Chern \: $= -2$) \cite{chen2020tunable}.


\subsection{Local-Sensing of Topological Invariants}

\hspace{2mm} Although local probes do not have direct access to the quantized Hall conductance of topological insulating phases, these techniques are capable of an alternate, equivalent measurement method. Here, scanning SET and STS measurements use the Streda formula to measure the Chern number $\mathbb{C}$ of these phases. Streda's theorem states that the charge density $n$ of a topological insulator changes with magnetic field $B$ at a rate equal to its quantized Hall conductance $\sigma_{xy}$ \cite{streda1982theory}:

\begin{equation}
        \frac{dn}{dB} = \frac{\sigma_{xy}}{e} = \frac{\mathbb{C}}{\Phi_0}
\end{equation}

\noindent  where $\Phi_0$ is the fundamental flux quantum. Streda's formula (above) describes a universal property of topological insulators, regardless of their origin. Transport experiments measure $\mathbb{C}$ directly via Hall conductivity measurements. However, local probes cannot access these observables. Instead, one can track the insulator's density $n$ as a function of magnetic field $B$. The Chern number $\mathbb{C}$ is, thus, inversely proportional to the slope of the state's linear trajectory in a plot of ($n$, $B$).

\hspace{2mm} This local method may seem circuitous, but there are two distinct advantages of this approach. First, local probes are excellent at identifying topological phases that are difficult to stabilize uniformly over micron-scale devices, but can be found in the highest quality nanometer-scale regions of a device. Second, local probes provide complementary observables that enable the unequivocal classification of these phases as correlation-driven Chern insulators. Alternatively, a newly developed nano-SQUID technique is also capable of local Chern number measurements by sensing the local change in magnetization as a function of chemical potential $dM/d\mu$, even at zero magnetic field \cite{grover2022chern}. In summary, these techniques provided a robust way to use local probe measurement schemes to access the topological invariant of the system.


\subsection{Correlation-Driven Topological Phases in MATBG}

\hspace{2mm} Correlation effects unlock previously inaccessible sources of Berry curvatures in the topological band crossings of 2D materials, producing correlated Chern insulators (CCIs) that break T-symmetry (Fig. \ref{fig:Topo}b) and symmetry-broken Chern insulators (SBCIs) that break both T-symmetry and moir{\'e}-translation symmetry (Fig. \ref{fig:Topo}c). Both are intrinsic to MATBG, which hosts a hierarchy of CCIs that emanate from every integer filling \filling \: $= \pm 1$, $\pm 2$, and $\pm 3$ with Chern numbers \Chern \: $= \pm 3$, $\pm 2$, and $\pm 1$ (Fig. \ref{fig:PhaseDiagrams}b). Many groups have observed these CCIs using complementary techniques \cite{nuckolls2020strongly, wu2021chern, saito2021hofstadter, das2021symmetry, choi2021correlation, park2021flavour, yu2022correlated}. Additionally, MATBG hosts SBCIs emanating from certain half-integer fillings \filling \: $= \pm 1/2$ and $\pm 3/2$ with Chern numbers \Chern \: $= \pm 3$ and $\pm 2$. These states have been observed in only a few devices to-date, and host energy gaps much smaller than the neighboring CCI states in MATBG \cite{saito2021hofstadter,yu2022correlated}.

\hspace{2mm} Scanning SET measurements of the compressibility of MATBG (Fig. \ref{fig:Topo}e) \cite{yu2022correlated} reveal a vibrant magnetic-field-dependent phase phenomenology resulting from many combinations of symmetry-breaking effects. By tracking the density of incompressible peaks as a function of magnetic field, scanning SET measures the Chern numbers of these insulators, uncovering the previously described sequence of CCIs (blue lines in Fig. \ref{fig:Topo}e) and a rare sighting of a \Chern \: $= -3$ SBCI from \filling \: $= -1/2$ (yellow line in Fig. \ref{fig:Topo}e). Similar approaches have now been used to discover topological phases in twisted TMD \moire materials \cite{kometter2022hofstadter,foutty2023mapping}.

\hspace{2mm} Importantly, negative compressibility features are a direct signature of correlation-driven phase transitions, and are seen near the incompressible peaks of each CCI. Some phase transitions were even observed to be hysteretic when sweeping either $n$ or $B$, indicating their first-order nature. Such measurements have uncovered crucial thermodynamic evidence for the phase transitions that give rise to CCIs in MATBG.

\hspace{2mm} Similar to SET experiments, STS measurements can track the densities of Chern gaps as a function of magnetic field to extract the Chern numbers of CCIs. In addition, Chern gaps in STS data appear to open and close at \EF, unassociated with single-particle gap features away from \EF \: (ex. \Chern \: $= -3$ CCI gap in MATBG; Fig. \ref{fig:Topo}f) \cite{nuckolls2020strongly, choi2021correlation}. Such spectroscopic behavior is an unequivocal signature that these CCIs result from electronic correlation effects. Hence, STS is an incisive probe of many-body topology. This technique has now been used to identify Chern phases in other \moire materials (ex. twisted monolayer-bilayer graphene) \cite{zhang2023local,zhang2022visualizing}.

\hspace{2mm} Correlation effects also produce fundamentally new kinds of topological phases without single-particle analogues, including fractional Chern insulators (FCIs) measured in recent scanning SET experiments on MATBG / hBN \cite{xie2021fractional}. Electronic compressibility measurements confirm a quantum anomalous Hall state near \filling \: $= +3$ \cite{sharpe2019emergent, serlin2020intrinsic, pierce2021unconventional}, further confirming that the density range between \filling \: $= +3$ and $+4$ is well-described by a single, isolated Chern \Chern \: $= - 1$ subband. At low magnetic fields (B $<$ $5$ T), calculations show that the Berry curvature of this Chern subband is concentrated in a hot-spot near the $\Gamma$ point of the \moire Brillouin zone. This Berry curvature distribution favors charge density wave states (Fig. \ref{fig:Topo}a and green lines in Fig. \ref{fig:Topo}g) that break moir{\'e}-translation symmetry. At moderate magnetic fields (B $\gtrapprox$ $5$ T), calculations show a redistribution of Berry curvature that favors FCIs (Fig. \ref{fig:Topo}d and red lines in Fig. \ref{fig:Topo}g), which break T-symmetry and host non-trivial topological order. Between \filling \: $= +3$ and $+4$, FCIs are observed in MATBG / hBN with \Chern \: $= +1/3$ (from \filling \: $= +11/3$) and \Chern \: $= +2/3$ (from \filling \: $= +10/3$). Very recently, FCI states have been observed at zero magnetic field in two new \moire materials. These include \Chern \: $= -2/3$ and $-3/5$ FCIs in tMoTe$_2$ (Fig. \ref{fig:PhaseDiagrams}c) \cite{cai2023signatures,zeng2023integer, park2023observation, xu2023observation}, and \Chern \: $= 2/5$, $3/7$, $4/9$, $4/7$, $3/5$, and $2/3$ FCIs in rhombohedral pentalayer graphene aligned to hBN \cite{lu2023fractional}, all of which satisfy the relation \filling \: $=$ \Chern. Such zero-field FCIs closely parallel the field-stabilized FCIs in MATBG / hBN, all occurring at partial fillings of a single isolated Chern subband.

\hspace{2mm} Scanning SET measurements not only uncover FCI states in MATBG / hBN in these simple settings, but also in more complex settings with multiple relevant Chern subbands \cite{xie2021fractional}. A \Chern \: $= -4/3$ FCI state emanates from \filling \: $= -5/3$, likely a \Chern \: $= 1/3$ FCI state that occurs when electron doping a nearby \Chern \: $= -1$ insulator. In addition, a \Chern \: $= -8/5$ FCI state emanates from \filling \: $= 11/10$, likely a symmetry-broken \Chern \: $= 2/5$ FCI state that additionally quadruples the \moire unit cell. Finally, the most exotic of the states reported are FCI states where the denominators of their Chern numbers and their associated fillings are co-prime. This is the case for a \Chern \: $= -5/2$ FCI state emanating from \filling \: $= -1/5$, which figuratively exists ``beyond the Standard Model'', as it can neither be described as a moir{\'e}-translation-symmetry-preserving nor a symmetry-broken FCI state. Further theoretical studies are required to provide an explanation for this unexpected observation.

\section{\label{sec:Ferro}Orbital Ferromagnetic and \Moire Ferroelectric Phases}
\subsection{New Material Design Principles for 2D Ferromagnets and Ferroelectrics}

\setlength{\columnsep}{0pt}
\setlength{\intextsep}{0pt}
\begin{figure*}
    \centering
    \includegraphics[width=\textwidth]{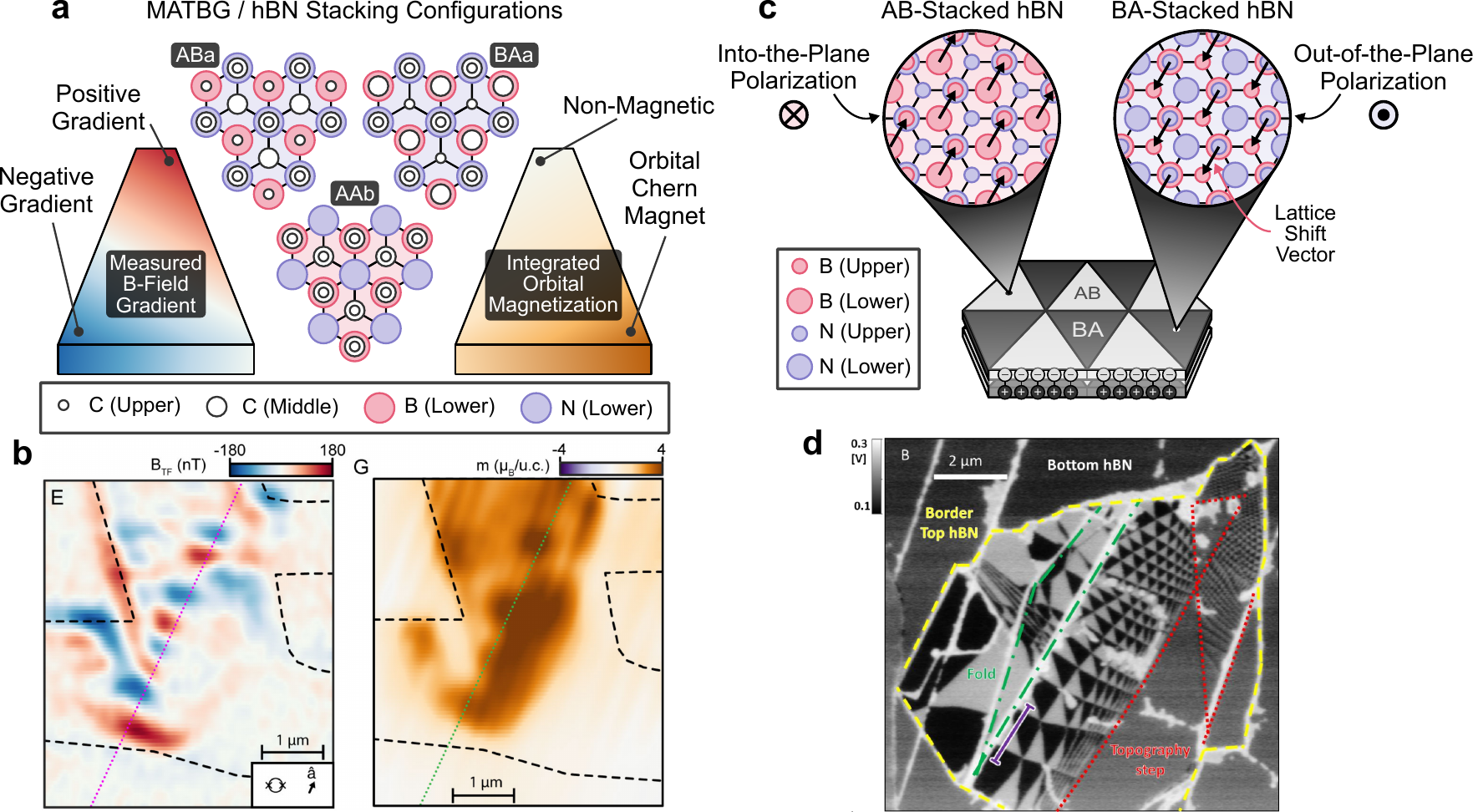}
    \caption{\label{fig:Ferro} \textbf{Imaging Orbital Ferromagnetism and \Moire Ferroelectricity.}  \textbf{a,} Schematic diagram of the local stacking configurations of MATBG commensurately aligned to hBN (center panel), as measured in STM imaging experiments. MATBG aligned to hBN hosts an orbital magnetic Chern insulator near \filling \: = $+3$ that breaks time-reversal symmetry spontaneously, as measured in scanning nano-SQUID measurements of the sample's field gradient (left panel schematic) and inferred orbital magnetization (right panel schematic). \textbf{b,} Scanning nano-SQUID measurements the magnetic-field gradient (left panel) and integrated orbital magnetization (right panel) of the quantum anomalous Hall state in MATBG aligned to hBN. \textbf{c,} Schematic diagram of twisted bilayer hBN, a room-temperature ferroelectric. This system purposefully breaks inversion symmetry, admitting two distinguishable, low-energy atomic stacking configurations that support a ferroelectric polarization pointing into-the-plane (-z direction; left panel) or out-of-plane (+z direction; right panel). \textbf{d,} Kelvin probe force microscopy (KPFM) image of twisted monolayer of hBN stacked atop a multilayer hBN flake. The local KPFM signal’s contrast (grey / black) distinguishes domains of different out-of-plane polarization. Reprinted from \cite{tschirhart2021imaging,vizner2021interfacial} with minor adaptations for style and formatting purposes.
    }
\end{figure*}

\hspace{2mm} \Moire materials research has introduced new paradigms in designing ferromagnetic and ferroelectric materials from 2D crystals that individually lack polarization. These new material design principles are a direct consequence of the unprecedented capabilities of van der Waals device fabrication techniques, which enable researchers to produce crystalline and \moire versions of materials that either purposefully or spontaneously break symmetries of their component 2D materials. In this section, we highlight recent studies using local electromagnetic probes that have provided distinct insights into the microscopic mechanisms of these novel orders.

\subsection{Imaging Orbital Ferromagnetism}

\hspace{2mm} Time-reversal symmetry breaking leads to ferromagnetism, of which there are broadly two types: spin and orbital ferromagnetism. The fundamental unit of magnetism in a spin ferromagnet is the spin of an electron, with an intrinsic magnetic dipole moment of $\mu_B = \frac{e \hbar}{2 m_e}$. Spin ferromagnets are common among materials containing atoms with unpaired spins, like iron, nickel, cobalt, and many rare-earth elements. Nearly all known ferromagnets are spin ferromagnets, deriving their macroscopic magnetization largely from the alignment of these spins (ex. 2D ferromagnets like CrX$_3$ (X = Cl, Br, I)). Spin-orbit coupling can cause a small amount of the material's spin ferromagnetism to mix into the orbital degree of freedom.

\hspace{2mm} In contrast, the fundamental unit of magnetism in an orbital ferromagnet is a spontaneous persistent current loop within the material's bulk, the dipole moment of which is material dependent. Orbital magnetization is exceedingly rare, and purely orbital ferromagnetic states (with negligible spin ferromagnetism) have only recently been discovered in \moire and crystalline 2D material devices \cite{sharpe2019emergent, serlin2020intrinsic, chen2020tunable, polshyn2020electrical, chen2021electrically, zhou2021half}. The macroscopic ferromagnetic response of an orbital ferromagnet can be understood using Green's theorem, according to which the internal circulation of orbital current loops can cancel in the bulk, but sums on the material's boundary to form a robust, chiral edge state - a topological edge state. A simple example of an orbital magnetic state is a Chern insulator, which will be the focus of this section. Even in \moire materials with negligible spin-orbit coupling (ex. MATBG), magnetic Chern insulators can still be stabilized by strong electron-electron interactions, which break T-symmetry spontaneously to leverage the intrinsic Berry curvature derived from the material's band structure, as discussed in Section \ref{sec:Topo}.

\hspace{2mm} The first \moire ferromagnet was observed in MATBG aligned to hBN (MATBG / hBN) \cite{sharpe2019emergent,serlin2020intrinsic}, where the system's atomic-scale stacking configurations were later identified by STM imaging experiments (center panels; Fig. \ref{fig:Ferro}a) \cite{oh2021evidence}. In this moir{\'e}-commensurate structure, the carbon-carbon ``AA'' regions of the MATBG \moire superlattice are aligned to the carbon-boron regions of the graphene-hBN \moire superlattice (labeled ``AAb''). The other two regions of the MATBG / hBN commensurate superlattice are inequivalent, stemming from the inequivalence of their corresponding graphene-hBN superlattice regions. ``ABa'' (``BAa'') regions form where atoms in the top (bottom) graphene sheet lie in register with atoms in the substrate's top hBN layer.

\hspace{2mm} Orbital ferromagnetism was first established by measurements of the anomalous Hall effect \cite{sharpe2019emergent,serlin2020intrinsic}, but later visualized directly using scanning nano-SQUID sensors \cite{tschirhart2021imaging}. Using a nanometer-scale SQUID device fabricated at the tip of a quartz pipette tip, local magnetic fields can be measured with unparalleled spatial resolution ($\approx 50$ nm). Raw magnetic-field-gradient measurements, obtained by oscillating the nano-SQUID tip with a tuning fork (left panel; Fig. \ref{fig:Ferro}b), reveal a rich orbital magnetic domain structure in MATBG / hBN. By subtracting images obtained within different branches of the Chern magnet's hysteresis loop to remove non-magnetic features from the data and by integrating the resulting gradient map, a map of the magnetization of the Chern state is obtained near \filling \: $= +3$ (right panel; Fig. \ref{fig:Ferro}b). 

\hspace{2mm} As a function of magnetic field, nano-SQUID measurements identify a direct correspondence between local magnetic domain configurations and the different anomalous Hall responses that they admit in transport measurements of the same device. These experiments give crucial insights into the distinctive switching mechanisms of orbital Chern states, as controlled by the sample's magnetic field environment. Similar nano-SQUID measurements have elucidated domain switching mechanisms in \moire heterobilayer TMDs, as controlled instead by currents passing through the sample \cite{tschirhart2023intrinsic}.

\subsection{Imaging \Moire Ferroelectricity}

\hspace{2mm} Ferroelectric / polar materials result from broken inversion symmetry, although not all inversion-broken materials are polar (ex. purely chirally stacked materials). Natural 2D ferroelectrics are rare, and often limited to those that admit purely in-plane polarization from in-plane inversion symmetry-breaking \cite{zhang2023ferroelectric}. Such in-plane ferroelectric materials can be difficult to integrate into future ferroelectric memory devices or ferroelectric tunnel junctions. 

\hspace{2mm} Although natural out-of-plane 2D ferroelectrics have been limited to crystals described by polar space groups, van der Waals stacking engineering techniques have recently enabled researchers to more generically combine inversion-symmetric 2D materials in stacks that purposely break the composite system's inversion symmetry \cite{yasuda2021stacking,vizner2021interfacial,woods2021charge,deb2022cumulative,kim2023electrostatic}. This can be done with perfect angle alignment, or with a slight angle misalignment that creates \moire materials with ferroelectric domains with opposite polarization directions derived from differences in their atomic-scale stacking configurations. Fig. \ref{fig:Ferro}c shows is a schematic of the local stacking configurations of ferroelectric domains in twisted bilayer hBN. Naturally exfoliated bilayer hBN exhibits an inversion-symmetric AA'-stacking, where each B (N) atom in the top layer lies in register with a N (B) atom in the bottom layer. Twisted bilayer hBN exhibits two low-energy stacking configurations, where AB- and BA-stacked domains appear with the same rotational orientation, but differ by an intra-unit-cell atomic displacement (Fig. \ref{fig:Ferro}c). Similar stacking approaches have been successful in producing new ferroelectric materials made from TMD layers as well \cite{wang2022interfacial,ko2023operando}.

\hspace{2mm} \Moire ferroelectrics have been detected in global transport experiments in devices that measure a proximate monolayer graphene sensor, which can be perturbed by doped charges induced by the polarization of twisted bilayer hBN \cite{yasuda2021stacking}. Contemporaneously, electric-field sensitive local probes like Kelvin probe force microscopy (KPFM; Fig. \ref{fig:Ferro}d) or electrostatic force microscopy (dc-EFM) have been used to directly visualize the electrostatic response of \moire ferroelectric phases \cite{vizner2021interfacial,woods2021charge}. Both techniques measure variations in a material's surface potential using a voltage-biased metallic tip, either through a direct measurement of the electrostatic force profile of the sample (dc-EFM) or via a PID-based feedback method that measures the voltage necessary to cancel the electrostatic force signal (KPFM).

\hspace{2mm} KPFM measurements of twisted multilayer hBN (Fig. \ref{fig:Ferro}d) provide direct evidence for a difference in electric polarization between the AB (grey shaded) and BA (black shaded) domains of the \moire superlattice. In addition, by scanning the tip across the sample with a large bias voltage, the ferroelectric domain structure can be manipulated on-demand by locally switching the polarization of a region of interest. Overall, these experiments demonstrate room-temperature ferroelectricity in stacking-engineered \moire ferroelectrics, which further show robust electrical switching dynamics that are crucial for next-generation non-volatile ferroelectric memory devices.

\section{\label{sec:SC}Unconventional Superconducting Phases}
\subsection{Low-Carrier-Density Superconductivity in \Moire Graphene}

\setlength{\columnsep}{0pt}
\setlength{\intextsep}{0pt}
\begin{figure*}
    \centering
    \includegraphics[width=\textwidth]{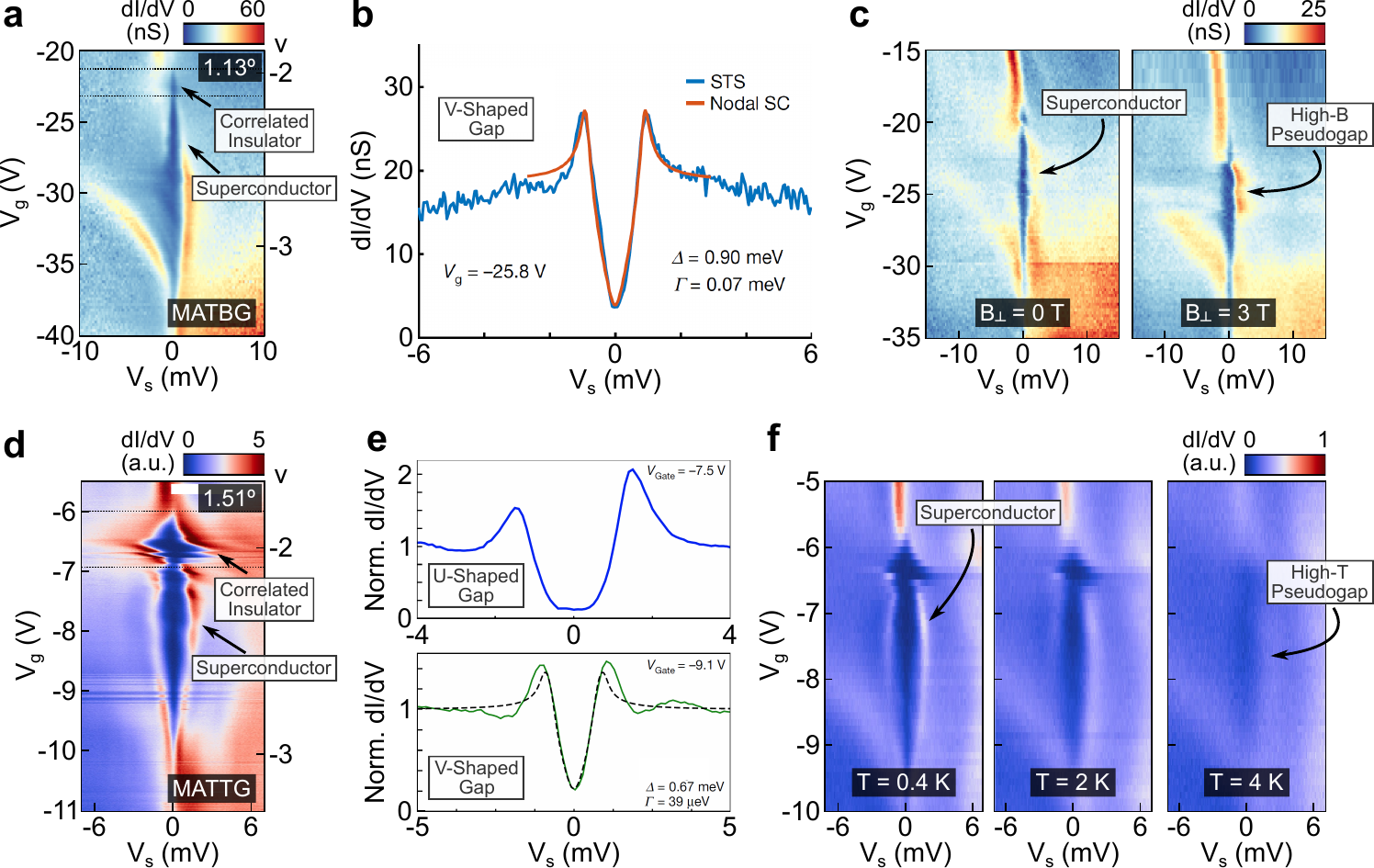}
    \caption{\label{fig:SC} \textbf{Spectroscopic Probes of Unconventional Superconductivity in \Moire Graphene.} \textbf{a,} Differential conductance \dIdVSG measured at the center of an AA site in magic-angle twisted bilayer graphene (MATBG). A gap at \filling \: = $-2$ (correlated insulator) opens and closes at the Fermi level, reopening into a gap between \filling \: = $-2$ and \filling \: = $-3$ (superconductor). \textbf{b,} dI/dV(V$_s$) spectrum obtained at V$_g$ = $-25.8$ V in \textbf{a}, fit to a nodal gap hypothesis (ex. p-wave, d-wave, f-wave). \textbf{c,} Differential conductance \dIdVSG measured as a function of magnetic field, showing the persistence of a spectroscopic gap beyond the critical magnetic field (B$_c$) of the superconducting phase in MATBG. This signature indicates a high-field pseudogap regime in MATBG. \textbf{d,} Differential conductance \dIdVSG measured near an AAA site in magic-angle twisted trilayer graphene (MATTG). \textbf{e,} dI/dV(V$_s$) spectra obtained at V$_g$ = $-7.5$ V (upper panel) The lower spectrum is fit to a nodal gap hypothesis (ex. p-wave, d-wave, f-wave). \textbf{f,} Differential conductance \dIdVSG measured as a function of temperature, showing the persistence of a spectroscopic gap beyond the critical temperature (T$_c$) of the superconducting phase in MATTG. This signature indicates a high-temperature pseudogap regime in MATTG. Reprinted from \cite{oh2021evidence,kim2022evidence} with minor adaptations for style and formatting purposes.
    }
\end{figure*}

\hspace{2mm} Transport measurements have uncovered superconducting phases in many graphene-based \moire materials \cite{cao2018unconventional, yankowitz2019tuning, lu2019superconductors, cao2021nematicity, park2021tunable, cao2021pauli, hao2021electric, park2022robust, zhang2022promotion, uri2023superconductivity, su2022superconductivity}, which occur at anomalously low-carrier densities at partial fillings of their flat bands. Superconductivity was first observed in MATBG, with a magic angle of $\theta_m \approx 1.1^\circ$ \cite{cao2018unconventional}, and subsequently confirmed in a series of alternating twisted $n$-layer graphene homostructures, where even-numbered layers are twisted to angle $\theta_n = 2\cos(\pi/(n+1)) \theta_m$ with respect to odd-numbered layers \cite{khalaf2019magic,park2021tunable, hao2021electric, park2022robust, zhang2022promotion}. All of these platforms show a sharp drop in resistance at their respective superconducting transition temperatures $T_c$, roughly $2$ - $3$ K in the best samples to-date, and show signatures of Fraunhofer oscillations far away from optimal doping, where superconductivity is weakest. Moreover, superconductivity in MATBG has been further confirmed in experiments on split-gate device geometries, yielding gate-tunable Josephson junctions that exhibit both DC and AC Josephson effects \cite{rodan2021highly, de2021gate, diez2023symmetry} and highly tunable superconducting quantum interference devices (SQUIDs) \cite{portoles2022tunable}, confirming the phase-coherence of this superconducting phase.

\hspace{2mm} Of particular interest is the idea that MATBG and MATTG appear to lie close to the Bose-Einstein Condensate (BEC) regime in the standard Uemura plot, which plots the ratio of a material's estimated Fermi temperature ($T_F$) to its measured critical temperature ($T_c$). The empirical similarities of the ratios of $T_c / T_F$ observed in twisted graphene multilayer superconductors to those of known correlated superconductors is highly suggestive of an unconventional pairing mechanism at play. However, conclusive evidence from transport measurements for a mechanism beyond the Bardeen-Cooper-Schrieffer (BCS) paradigm is absent, and is largely inaccessible due to the difficulty of performing global transport experiments that are phase-sensitive (i.e. sensitive to the phase $\phi (r)$ or the complex order parameter $\psi (r) = | \psi (r) | \: e^{i \phi (r)}$ in the Ginzburg-Landau theory of superconductivity).

\hspace{2mm} In contrast, spectroscopic probes, including scanning tunneling spectroscopy (STS) and point-contact spectroscopy (PCS), have access to the relevant physical observable for discerning the nature of superconductivity in twisted graphene multilayer superconductors. In this Section, we highlight recent STS and PCS measurements on MATBG and MATTG, which now provide a preponderance of evidence for unconventional, non-BCS superconductivity in these platforms \cite{oh2021evidence, kim2022evidence}. Moreover, spectroscopic probes have shown remarkable similarities between these two platforms, including agreement on certain qualitative features of their tunneling gap shapes, their Andreev reflection spectra, and the shared presence of a phase-incoherent pseudogap regime that persists above the critical temperature and critical magnetic field of their superconducting states.


\subsection{Debating the Nature of Superconductivity in \Moire Graphene}

\hspace{2mm} We briefly review what was understood prior to STS experiments about the nature of superconductivity in twisted multilayer graphene. Although we focus on such experiments performed on MATBG, the origins of superconductivity in MATBG and MATTG are expected to be related, and were the subject of immense debate, ever since their initial reports \cite{cao2018unconventional, park2021tunable, cao2021pauli, hao2021electric}. On one hand, several characteristics of superconductivity were highly suggestive of an unconventional pairing mechanism. In both materials, superconductivity appeared at anomalously low carrier densities, which presented an apparent mismatch of energy scales for BCS theory to be applicable. Superconductivity also appears in a phase diagram broadly similar to that of the high-temperature cuprate superconductors, where an insulator appeared at \filling \: $= -2$, flanked by superconducting domes when either electron- or hole-doped \cite{sigrist1991phenomenological, tsuei2000pairing, lee2006doping, fradkin2015colloquium, keimer2015quantum}. Unconventional superconductivity appears proximate to a correlated insulator not only in cuprates superconductors, but also in the iron-pnictide superconductors \cite{johnston2010puzzle, wen2011materials, stewart2011superconductivity, si2016high, fernandes2022iron}, heavy fermion superconductors \cite{stewart1984heavy, petrovic2001heavy, white2015unconventional}, and organic superconductors \cite{jerome1991physics, singleton2002quasi}.

\hspace{2mm} On the other hand, further transport studies showed that superconductivity in MATBG might be more conventional than was initially suggested by measuring superconductivity in MATBG when placed in close proximity to another two-dimensional material (often a metal). Superconductivity in MATBG was at least persistent \cite{stepanov2020untying, arora2020superconductivity, saito2020independent}, if not more robust in the presence of screening \cite{liu2021tuning}. The critical magnetic field, the critical temperature, the critical current, and the density range over which superconductivity were all observed to systematically increase by a few percent in the presence of electronic screening. Such data, however, has several plausible interpretations. Electronic screening could suppress correlated phases that compete with conventional superconducting states, but it could also protect unconventional superconducting states from the influence of defect scattering \cite{anderson1959theory,zeljkovic2013interplay}.

\hspace{2mm} In addition, early theoretical work had described many possible superconducting pairing mechanisms \cite{xu2018topological, wu2018theory, po2018origin, isobe2018unconventional, liu2018chiral, gonzalez2019kohn, guinea2018electrostatic, xie2020topology, kennes2018strong, peltonen2018mean}, some of which are purely conventional (ex. singlet pairs mediated by acoustic phonons \cite{lian2019twisted}) while other proposals appeared singularly unconventional (ex. topological superconductivity resulting from charged isospin skyrmion pairing \cite{khalaf2021charged}).

\hspace{2mm} Although transport experiments clarified the competitive relationship between the correlated insulating and superconducting orders in MATBG, their ability to discern the nature of superconductivity itself is indirect and often subject to interpretation. In contrast, the energy resolution of spectroscopic probes enables the identification of signatures of unconventional superconductivity, many that are irreconcilable with the conventional BCS theory \cite{damascelli2003angle, fischer2007scanning}. Signatures of a superconducting state's unconventional pairing symmetry or pairing mechanism provide crucial constraints for constructing successful theories of these exotic phases. Such information was the subject of recent spectroscopic probes of superconductivity in MATBG and MATTG.


\subsection{Characterizing Superconductivity in MATBG and MATTG}

\hspace{2mm} One key, shared observation in STS measurements is a nodal tunneling spectrum in the superconducting states of MATBG (Fig. \ref{fig:SC}b) and MATTG (Fig. \ref{fig:SC}e; lower panel) \cite{oh2021evidence,kim2022evidence}. In contrast to the s-wave tunneling gaps observed in prototypical BCS superconductors (ex. Al, Zn, Sn), as characterized by sharp coherence peaks at the gap edge that flank a zero-conductance ``hard-bottom'' spectroscopic gap, tunneling gaps in MATBG and MATTG (at some densities) appear to have broader coherence peaks with linearly sloped gap edges that join in a cusp-like, nodal minimum at zero-bias.

\hspace{2mm} In addition, MATTG shows a transition in its spectroscopic properties, not seen in MATBG. In MATTG, a U-shaped gap appears at densities closest to \filling \: $= -2$ (Fig. \ref{fig:SC}e; upper panel), which transitions into a V-shaped gap upon hole-doping the system \cite{kim2022evidence}. This transition is suggestive of either a transition in the pairing symmetry of the superconducting state, or in the coupling regime (i.e. BEC-BCS) for superconductivity dictated by a single (nodal) pairing symmetry. Most importantly, although gaps in both platforms vary from sample to sample, none of the reported tunneling gaps resemble the isotropic s-wave prediction of BCS theory.

\hspace{2mm} In addition, the tunneling gaps of the superconducting states in MATBG and MATTG are generically much larger than expected for $T_c \approx 2-3$K. This feature is shared among correlated superconductors (ex. cuprate and iron-based superconductors), quantified by the ``gap-to-$T_c$'' ratio ($2 \Delta / k_B T_c$). In BCS theory, the size of the superconducting gap $\Delta (T = 0) \equiv \Delta$ is proportional to $T_c$ with a universal proportionality constant $2 \Delta / k_B T_c \approx 3.528$. In many conventional BCS superconductors (ex. Al), this universal ratio is reflected experimentally to a remarkable degree. In superconductors with strong electron-phonon coupling, Eliashberg theory predicts higher estimates of this ratio (ex. $5$ - $6$ for some alloys of Pb and Bi). In contrast, gap-to-$T_c$ ratios for MATBG and MATTG appear as large as $15$ - $25$, which underscores the need to be cautious when interpreting these spectra.

\hspace{2mm} Although gaps in MATBG and MATTG often show good agreement with a simple, nodal gap function, the large gap size could suggest these tunneling gaps to be instead attributed to a pseudogap state. Drawing from STS studies of the cuprates \cite{fischer2007scanning}, nodal gap fits of the low-bias region of spectra of MATBG were performed to show that the low-energy density of states is best described by a nodal hypothesis for every density throughout MATBG's superconducting dome. Following this work came thermal conductance measurements of the superconducting state of MATBG, supporting this interpretation by demonstrating a temperature-dependence consistent with a nodal superconducting gap \cite{di2022revealing}.

\subsection{High-Temperature and High-Field Pseudogap Regimes}

\hspace{2mm} Finally, STS measurements of MATBG and MATTG both uncover a precursor pseudogap regime, which persists beyond the $T_c$ and $H_c$ of superconductivity. Both studies used point-contact spectroscopy to measure the presence or absence of superconductivity locally, which unequivocally distinguishes these states by the presence or absence of an Andreev reflection spectrum \cite{deutscher2005andreev}. Although the Andreev spectrum vanishes near $T_c$ ($\approx 1$ - $2$ K), a deep suppression of the density of states near \EF \: persists to T $>$ $5$ K ($>$ $6$ K) in MATBG (MATTG) (Fig. \ref{fig:SC}f), indicative of a high-temperature pseudogap regime from which superconductivity is derived at base temperature.

\hspace{2mm} Moreover, although the Andreev spectrum vanishes near $H_c$ ($\approx 100$ mT), a deep suppression of the density of states near \EF \: persists to B $>$ $6$ T ($>$ $4$ T) in MATBG (MATTG) (Fig. \ref{fig:SC}c), indicative of a high-field pseudogap regime from which superconductivity is derived at zero magnetic field. Overall, these pseudogap regimes in MATBG and MATTG provide additional evidence for unconventional superconducting states in these material, for the BCS gap vanishes when superconductivity is quenched.

\hspace{2mm} There is still much to be understood about the nature of these unconventional superconducting states. Looking forward, we anticipate that future STM studies will perform quasiparticle interference experiments near shallow defects to further elucidate the pairing symmetry of these states, and will perform scanning Josephson spectroscopy experiments to probe the coupling between superconducting and isospin-symmetry-breaking order parameters.

\section{\label{sec:Outlook}Outlook}
\hspace{2mm} The distinctive tunability of \moire materials has enabled high-spatial-resolution studies of their complex symmetry-broken phases. While local probe experiments have mainly focused on a small subset of \moire materials, transport and optical studies are rapidly uncovering new correlated and topological phases in this developing class of materials. Surprisingly, each new \moire material has presented a unique facet of many-body physics, with new electronic phases not seen in other \moire materials. This trend is likely a reflection of the subtle interplay of many relevant energy scales (ex. on-site Coulomb energy, spin-orbit coupling) that affect correlated electronic states in these materials. As described in this Review, the microscopic origins of these phases are best characterized by physical observables accessible to local probes, a perspective that will shape our future understanding of these exotic quantum phases.

\hspace{2mm} Going forward, we believe that there are three research focuses that are particularly important to the near-term goals of this community. First, our community will benefit from addressing several technical limitations of existing local probes that impede our ability to better understand \moire materials. This includes a focus on new techniques for preserving air-sensitive \moire materials (ex. t\MoTe) in ways compatible with surface-sensitive local probes (ex. STM, KPFM), and new methods for conducting local electromagnetic measurements (ex. nano-SQUID, MIM) at millikelvin temperatures required for stabilizing fragile correlated phases.

\hspace{2mm} Second, our community will benefit from further developing new local probe techniques introduced in the \moire material era. Atomic-scale charge-sensing experiments using the STM, as used to image generalized Wigner crystal states in \WSe / \WS \cite{li2021imaging}, have great potential in future studies of \moire materials, and might even be capable of quantitative measurements of related physical observables (ex. electronic compressibility) at the nanometer scale. A new nano-SQUID technique can image thermodynamic quantum oscillations in 2D devices to reconstruct local band structure features, which has the potential to provide sensitive measurements of subtle \moire flat band features \cite{zhou2023imaging,bocarsly2024haas}. The quantum twisting microscope, a revolutionary new measurement tool capable of local 2D wavefunction interference experiments using van der Waals device-structured sensors, has opened up new opportunities to manipulate \moire materials on demand \cite{inbar2023quantum}. Further developing this tool's momentum-space imaging capabilities, and exploring new applications of its novel sensor geometry will support an important new microscopic perspective on \moire materials.

\hspace{2mm} Finally, our community will benefit from continuing to search in \moire materials for exotic quantum phases unrealized elsewhere in Nature, leveraging the many advantages of local probe techniques. Nearly all electronic phases observed in \moire materials to-date have been studied to some degree in other material classes, albeit \moire materials offer highly tunable platforms that enable new investigative approaches. However, we emphasize the importance of finding and understanding novel phases without other material realizations, like the unique time-reversal breaking superconducting state found in twisted cuprate interfaces \cite{zhao2023time} and the non-equilibrium ferromagnetic and bosonic correlated insulating states found in \WSe / \WS \cite{wang2022light, xiong2023correlated}. Such efforts will support the greatest impact of this budding research field. An exciting prospect is the recent identification of fractional Chern insulators (FCIs) stabilized at zero magnetic field. FCIs have now been reported in t\MoTe \cite{cai2023signatures,zeng2023integer, park2023observation, xu2023observation} and in rhombohedral pentalayer graphene aligned to hBN \cite{lu2023fractional}, and very recent theoretical studies have suggested that such phases may be stabilized in other \moire materials \cite{crepel2023chiral,dong2023anomalous,dong2023theory,kwan2023moir,reddy2023fractional,zhou2023fractional,morales2023magic, dong2024stability}.

\hspace{2mm} Understanding a series of open questions about the nature of FCI states will be a major research focus of local probe experiments in the coming years. To what extent do the recently observed Jain-type FCIs (i.e. \Chern \: $= \pm 2/3$, $ \pm 3/5$ FCIs) parallel Jain states observed in the fractional quantum Hall regime? What kinds of zero-field fractional phases can we realize that are fundamentally different from those realized at high magnetic fields (ex. time-reversal symmetric fractional states, new non-abelian states) \cite{kang2024observation,reddy2024non,fujimoto2024higher,may2024theory}? How can we understand FCI states ``beyond the Standard Model'', which have no clear fractional quantum Hall analogues (ex. \Chern \: $= -5/2$ FCI emanating from \filling \: $= -1/5$ in MATBG / hBN)? Key to answering these questions will be the microscopic perspective of local probe techniques, which access distinguishable physical observables of these complex ground states and can surpass the limitations of sample inhomogeneity to identify fragile FCI phases in the cleanest, nanometer-scale domains of \moire material devices. Furthermore, certain local probe techniques compatible with top-gated device geometries, including scanning nano-SQUID and atomic-scale charge-sensing using the STM, are distinctly capable of accessing the complete phase diagrams of t\MoTe and rhombohedral graphene aligned to hBN, which we expect will provide early microscopic insights into these FCIs.

\centering
{\bf Acknowledgements}

\raggedright
We are grateful to all of our collaborators on this subject, particularly to D. Wong, M. Oh, R. L. Lee, T. Soejima, J. P. Hong, D. C{\u{a}}lug{\u{a}}ru, J. Herzog-Arbeitman, C. Chen, Y. Xie, B. Lian, Y. Chen, O. Vafek, N. Regnault, M. Zaletel, and B. A. Bernevig. In addition, we thank J. G. Checkelsky for fruitful discussions during the writing of this Review. K. P. Nuckolls acknowledges support from the MIT Pappalardo Fellowship in Physics. A. Yazdani acknowledges funding from Gordon and Betty Moore Foundation’s EPiQS initiative grant GBMF9469, DOE-BES grant DE-FG02-07ER46419, ONR grant N00014-21-1-2592, NSF-MRSEC through the Princeton Center for Complex Materials grant NSFDMR-2011750, the U.S. Army Research Office MURI project under grant number W911NF-21-2-0147, and NSF grant DMR-2312311.


\bibliography{refs}

\end{document}